\PassOptionsToPackage{table,dvipsnames}{xcolor}
\documentclass[sigconf]{acmart}

\AtBeginDocument{%
  \providecommand\BibTeX{{%
    \normalfont B\kern-0.5em{\scshape i\kern-0.25em b}\kern-0.8em\TeX}}}

\copyrightyear{2026}
\acmYear{2026}
\setcopyright{cc}
\setcctype{by}
\acmConference[CHI '26]{Proceedings of the 2026 CHI Conference on Human Factors in Computing Systems}{April 13--17, 2026}{Barcelona, Spain}
\acmBooktitle{Proceedings of the 2026 CHI Conference on Human Factors in Computing Systems (CHI '26), April 13--17, 2026, Barcelona, Spain}
\acmPrice{}
\acmDOI{10.1145/3772318.3791518}
\acmISBN{979-8-4007-2278-3/2026/04}

\usepackage{graphicx} 
\usepackage{tabularray}
\usepackage{multicol}
\usepackage{booktabs}
\usepackage{appendix}
\usepackage{caption}
\usepackage{subcaption}
\usepackage{array}
\usepackage{tabularx}
\usepackage{enumitem}
\usepackage{colortbl}

\begin{document}

\title{Fostering Collective Discourse: A Distributed Role-Based Approach to Online News Commenting}

\author{Yoojin Hong}
\email{dbwk18@kaist.ac.kr}
\affiliation{
  \institution{KAIST}
  \streetaddress{291 Daehak-ro, Yuseong-gu}
  \city{Daejeon}
  \country{South Korea}
  \postcode{34141}
}

\author{Yersultan Doszhan}
\email{edoszhan@kaist.ac.kr}
\affiliation{
  \institution{KAIST}
  \streetaddress{291 Daehak-ro, Yuseong-gu}
  \city{Daejeon}
  \country{South Korea}
  \postcode{34141}
}

\author{Joseph Seering}
\email{seering@kaist.ac.kr}
\affiliation{
  \institution{KAIST}
  \streetaddress{291 Daehak-ro, Yuseong-gu}
  \city{Daejeon}
  \country{South Korea}
  \postcode{34141}
}

\begin{abstract}
  Current news commenting systems are designed based on implicitly individualistic assumptions, where discussion is the result of a series of disconnected opinions.
This often results in fragmented and polarized conversations that fail to represent the spectrum of public discourse.  
In this work, we develop a news commenting system where users take on distributed roles to collaboratively structure the comments to encourage a connected, balanced discussion space. 
Through a within-subject, mixed-methods evaluation (N=38), we find that the system supported three stages of participation: understanding issues, collaboratively structuring comments, and building a discussion. 
With our system, users' comments displayed more balanced perspectives and a more emotionally neutral argumentation. Simultaneously, we observed reduced argument strength compared to a traditional commenting system, indicating a trade-off between inclusivity and depth. 
We conclude with design considerations and trade-offs for introducing distributed roles in news commenting system design.
\end{abstract}




\maketitle

\section{Introduction}

The expansion of the internet has transformed commenting sections in news outlets into essential platforms for public discourse. Initially, these sections were seen as spaces where individuals from diverse backgrounds could share their opinions on equal footing, fostering a rich and inclusive exchange of ideas. While the expectation was that these platforms would democratize the conversation, allowing for a variety of perspectives to be heard, this ideal has often not been realized~\cite{ksiazek2015discussing, ruiz2011public}. Instead of being spaces for constructive discussion, they have frequently become spaces for polarized~\cite{hsueh2015leave} and uncivil exchanges~\cite{coe2014online}. This failure has led many news outlets to shut down their commenting sections~\cite{nelson2021killing}, thereby limiting public engagement in online discussions.

Previous studies have explored how the design of commenting systems can shape user behavior~\cite{kiskola2021applying, seering2019designing}. In many current news outlets, standardized commenting systems are used, where users simply post their comments or reply to others with limited opportunities to interact with other participants. This design often results in comments that are isolated expressions of personal thoughts, making it difficult for meaningful discussions to develop. As a result, discussions tend to unfold as a series of individual expressions, with users primarily focused on sharing their own views rather than collaborating toward a collective understanding or outcome. This behavior leads to fragmented conversations~\cite{dahlberg2007rethinking}, where the overall discussion lacks depth. Consequently, similar viewpoints are often amplified, while diverse perspectives remain underrepresented~\cite{anderson2014nasty}.

In this paper, we aim to redesign the current online commenting system to move beyond the mere expression of opinions toward actively shaping a more constructive dialogue, where users can interact with diverse perspectives, expand their views, and contribute to discussions that not only reflect their own values but are also informed by the complexities of different viewpoints. 
To achieve this, our core idea is to implement ``distributed roles’' within the discussion platform, enabling users to not only contribute their opinions but also have specific roles to engage in organizing the conversation by integrating and reviewing others' comments. This approach is intended to enrich the discussion by increasing comprehension of others' viewpoints through the process, while also fostering a sense of community and shared responsibility for thoughtful contributions to the discussion space.

Our system introduces three main features~--~clustering, summarizing, and threading~--~collaboratively organized by distinct user roles. \textbf{Clustering} groups comments with similar themes, encouraging users to consider different perspectives and follow coherent threads of conversation. \textbf{Summarizing} consolidates the clusters into cohesive summaries, reducing redundancy and highlighting key points. \textbf{Threading} organizes comments into subtopics, fostering deeper engagement and focused dialogue on various aspects of the issue at hand. Together, these roles contribute to a discussion space that fosters collective understanding by addressing complex viewpoints in a structured manner. Our system was developed as a browser extension to ensure adaptability and compatibility with most news outlet websites. 

To evaluate how our system aids in collective understanding compared with the conventional commenting interface, we conducted a within-subjects study with 38 participants. Each of the participants was pre-assigned to a specific user level based on our study configuration, which remained consistent throughout the experiment. We conducted a post-survey after using our system, followed by interviews with 14 participants who were willing to take part. 
Specifically, we ask the following research questions:
\begin{itemize}
    \item \textbf{RQ1.} How does a structured discussion system influence patterns of user engagement?
    \item \textbf{RQ2.} In what ways does the system affect the quality of user comments?
    \item \textbf{RQ3.} How does the system support users’ experiences of participating in online discussions?
\end{itemize}

Our study's quantitative findings revealed that the system increased the overall volume of comments while producing significantly shorter contributions, indicating a shift toward more frequent and concise participation. We present detailed statistics on system usage and the outputs it generated, highlighting the dynamics of tension involved in performing assigned activities and their role in deliberately shaping the discussion space. An analysis of the comments indicates that the system fostered a more balanced distribution of perspectives but did not significantly expand the range of distinct viewpoints. At the same time, comments exhibited reduced argumentative support and lower levels of emotional expression, while politeness remained unchanged. Taken together, these findings suggest that the system promoted balanced and neutral participation, accompanied by a shift away from emotionally charged and heavily reasoned styles toward clearer, more focused forms of engagement. 

The finding also highlights improvements in users' understanding of the issue, the flow of discussion, and awareness of diverse perspectives. We found that this enhancement was achieved through the system's impact on key stages of the commenting process, including 1) reading the article and comprehending the ongoing discussion, 2) structuring thoughts and writing comments, and 3) contributing to the discussion. We explored how the system facilitated these stages and identified recurring themes that highlight its contributions to each phase. Additionally, we addressed aspects that could potentially limit participation, discussing both concerns and the benefits of our system's features. The findings of our study demonstrate the value of distributed roles in fostering constructive discussion, promoting thoughtful and engaging participation while respecting diverse perspectives. We conclude by discussing the implications of our system and exploring the design space for building collaborative discourse. 

Our research contributes to the exploration of an untapped space in designing collective online discourse by introducing a collaborative approach centered on ``distributed roles,'' reimagining how commenters can collectively shape and moderate discussions. We design and implement a novel system grounded in insights from previous works. Additionally, through experimentation, we present detailed findings on how our system differently shapes user behavior, understanding, and identifying the newly emerged behavioral patterns using our suggested system. Finally, through our design exploration, we identify key design considerations and trade-offs associated with the values embedded in our system, which can serve as guidance for future efforts in designing collective discourse systems.

\section{Related Work}

\subsection{News Commenting as a Public Sphere 2.0}
With the rise of the internet, news outlets have become crucial in shaping the public sphere. The deliberation of diverse views has always been essential for building strong democratic societies~\cite{habermas1991structural, dewey2012public}. The internet's horizontal structure connected diverse individuals in both one-to-one and also many-to-many conversations, giving more people a public voice and enabling public debate. In response, news outlets have started creating infrastructures that support public discussion of the news, and the commenting sections were adopted by most of the top 150 U.S. newspapers~\cite{santana2011online}.

By inviting reader comments, today’s news media have embraced greater user involvement in the journalistic process. This phenomenon has been explored by scholars through frameworks such as ``participatory journalism''~\cite{singer2011participatory} and ``reciprocal journalism''~\cite{lewis2014reciprocal}; the audience has also become an active contributor to content. While there were positive views of news commenting acting as the digital cafés of a Public Sphere 2.0~\cite{ruiz2011public}, journalists and news outlets were often concerned about quality control, manageability, and the maintenance costs of user-generated content~\cite{diakopoulos2011towards}.

User comments have increasingly become an attractive playground for dark participation~\cite{quandt2018dark}, resulting in a surge of problematic behaviors, including hate speech~\cite{erjavec2012you, hughey2013racist}, incivility~\cite{coe2014online, muddiman2017news}, trolling~\cite{wolfgang2021taming}, and flaming~\cite{o2003reconceptualizing}. News organizations had to develop norms and practices to combat these problematic behaviors, such as requiring to use their real names~\cite{reich2011user}, allowing commenters to flag comments~\cite{paulussen2011inside}, and removing comments that don’t meet journalistic standards~\cite{mcelroy2013old}. Despite these efforts, continuing concerns about harming their brand and the resources needed to monitor and moderate comments daily~\cite{reimer2023content} have led an increasing number of news outlets to shut down their comment sections.

Overall, the literature on online news comments indicates that journalists are skeptical about the quality of audience contributions in news forums~\cite{reich2011user, robinson2010traditionalists}. As a result, they often chose to limit the extent of user participation in the process. To help preserve the role of user comments as a public sphere, our work proposes a way to improve the quality of audience contributions without requiring post-managing efforts from news outlets or limiting user participation. Through this, we aim to address the challenge news outlets face in moderating and shaping their commenting spaces into something resembling the idealized dialogue of the public sphere.

\subsection{Improving Discussion Quality through Moderation} 
Moderation systems in online commenting platforms are crucial for enhancing discourse quality, and various strategies have been developed to manage and improve interactions. These approaches range from direct intervention by moderators to more community-driven methods. Broadly, two main attitudes toward moderation have emerged: pre-moderation, which is a more interventionist approach, and post-moderation, which is a more relaxed approach~\cite{reich2011user}. 

A pre-moderation approach seeks to improve the discussion quality by addressing potential issues before they arise. These proactive approaches range from using design elements to nudge users toward prosocial participation~\cite{seering2019designing}, to more direct interventions that review and approve comments before they appear publicly. Common design elements include explicitly displaying community rules and guidelines~\cite{kraut2012building, matias2019preventing}, using prompts to encourage more thoughtful participation~\cite{mcinnis2016one}, and employing curation strategies that influence users' deliberation~\cite{mcinnis2018effects}. More direct strategies involve using algorithmic support to provide risk information to users~\cite{chang2022thread} or moderators~\cite{schluger2022proactive}, helping in the identification and resolution of tensions before they escalate.

The post-moderation approach, by contrast, takes place after content has been published. It focuses on identifying and mitigating the impact of harmful or inappropriate material through reactive review of comments. While traditional methods involve human moderators reviewing flagged content~\cite{roberts2019behind, seering2019moderator}, the scale and volume of content necessitates heavy reliance on community participation to flag or report inappropriate comments. Community-driven methods, also called a crowd-based approach, have proven to be effective for filtering and evaluating the quality of comments~\cite{diakopoulos2011towards}.
This approach leverages the collective judgment of the community to assess and organize comments, fostering a more dynamic and responsive environment. Collaborative platforms such as Wikipedia exemplify this, where community members work together to resolve disputes~\cite{im2018deliberation} and contribute to the creation of high-quality articles~\cite{zhang2017crowd}. Another notable example is the moderation system used by Slashdot~\cite{lampe2007follow}, where users rate and moderate comments, with these decisions aggregated to determine comment quality. Slashdot's effectiveness lies in its ability to harness the collective intelligence of its users, allowing for real-time participation in organizing and structuring discussions. This distributed form of moderation is both scalable and adaptable, addressing the diverse needs of its community. 

Our proposed system builds on the principles of crowd-based moderation seen in Slashdot by further enhancing participation quality through distributed roles within the discussion. By encouraging users to take on specific responsibilities in contributing to discussions, our system aims to improve the overall quality of discourse. This approach not only benefits real-time participation but also fosters a sense of agency and engagement among users, ensuring that they are more invested in maintaining high standards within the community.

\subsection{Restructuring and Reimagining the Discussion Space}
The deliberative model of democracy emphasizes the importance of reasoned argumentation, mutual respect, and the willingness to consider others’ perspectives as essential components for effective discourse~\cite{dahlberg2007internet}. This framework has informed a variety of approaches to structuring online discussions, with the goal of enhancing the quality and inclusiveness of these spaces. Researchers have explored multiple avenues for improving discussion forums, including the development of structured workflows~\cite{farnham2000structured, kim2021moderator, lee2020solutionchat}, the integration of structural redesigns that facilitate more nuanced conversation~\cite{zhang2017wikum, tian2021system, nam2007arkose, rambow2004summarizing, zhang2018making}, and features that specifically facilitate the promotion of diverse viewpoints~\cite{faridani2010opinion, munson2013encouraging, kriplean2012supporting, kim2021starrythoughts}. These efforts aim to create more organized, engaging, and meaningful discussions by combining the strengths of community-driven processes and automated tools.

Previous studies have introduced various structural design changes aimed at improving navigation of and participation in long-tailed discussions. These structural representations included clustering, summaries, and threads to easily navigate and gain overview of the discussion. Clusters and threads have been found to be useful to identify insightful comments while navigating complex structures of discussion at varying levels of detail. For example, ordering comments into visual clusters~\cite{hoque2015convisit} and hierarchical thematic organization~\cite{hoque2016multiconvis} supported users in identifying insightful contributions and easily narrowing down to a subset of conversations. Adding a thread structure in a discussion space has been shown to increased user retention in comment participation~\cite{budak2017threading}, showing how structural design changes can also impact the user behavior. Summaries have played a crucial role in the synthesis of ideas within the discussion to encourage participants to reflect on the conversation~\cite{kriplean2012you} and provide a structured overview that help readers navigate the main topics~\cite{zhang2017wikum}. Systems like Wikum~\cite{zhang2017wikum} exemplify this approach by employing recursive summarization, enabling users to distill key points from extensive discussions. This method expanded to the creation of ``living summaries''~\cite{tian2021system} that evolve as new contributions and insights emerge, ensuring that the discourse remains coherent and accessible. 
Beyond these approaches, some efforts have focused on developing lightweight tools to add structure that support easier contribution in the unstructured discussions spaces~\cite{schneider2011understanding, zhang2018making}. While these techniques have improved the accessibility of information by restructuring the discussion, their potential to actively foster the collaborative building of shared understanding and collective insights during the discussion process remains underexplored.

Beyond structuring workflows, another crucial aspect of improving discussion spaces is promoting the inclusion of diverse viewpoints. Encouraging exposure to a variety of perspectives not only enriches the conversation but also helps users recognize and challenge their biases. Tools like Balancer~\cite{munson2013encouraging} and ConsiderIt~\cite{kriplean2012supporting} are designed to nudge users toward engaging with content from different sources and perspectives, fostering a more balanced and informed discourse. By actively recommending opposing viewpoints, systems like those developed by Gao et al.~\cite{gao2018burst} and Nelimarkka et al.~\cite{nelimarkka2019re} mitigate selective exposure, encouraging users to explore opinions that differ from their own. 
Systems such as Reflect~\cite{kriplean2012you} promote active listening by summarizing others’ comments, leading to a deeper understanding of the intention of the commenter. In addition, interacting with moral framing grounded in frameworks such as Moral Foundations Theory (MFT) enables users to reflect on shared moral values underlying different perspectives, encouraging to rethink and engage with opposing viewpoints~\cite{wang2022designing}. Various visualization methods have also been used to map the diversity of users' opinions and help navigate them effectively~\cite{faridani2010opinion, kim2021starrythoughts}.

These two strands of research~--~structured workflows with different structural representations and the promotion of diverse viewpoints~--~are interconnected in their aim to create more constructive and inclusive online discussion platforms. While structured workflows ensure that discussions remain organized and focused, the inclusion of diverse perspectives ensures that the conversation is rich and multifaceted. Our system design builds on these foundations by integrating structured workflows that organize the discussions by user-driven activities to capture and synthesize the evolving conversation. Simultaneously, our approach actively promotes the inclusion of diverse perspectives to ensure that a wide range of viewpoints is represented and engaged throughout the organized discussion. 

Together, these approaches advance our understanding of how to design discussion spaces that are well-organized and also enriched by the diversity of thought, thereby reimagining the potential of online discourse.

\section{System Design and Implementation}

In this section, we outline our design approach and system implementation, beginning with the inspiration drawn from the community-driven moderation model of Slashdot. We then explain the role assignment process, detailing how each level of user participation contributes to the collective organization and development of the discussion space. Following an overview of implementation details, we present the core features of our system~--~clustering, summarizing, and threading~--~and describe how these features are achieved through a distributed system of user roles.

\subsection{Design Inspiration from `SlashDot'}
Our system is inspired by Slashdot\footnote{https://slashdot.org/}, a platform known for its distributed approach to managing user behaviors in large-scale online discussions. Slashdot’s system allows users to take on specific roles in moderating the comments, with three levels of user participation: moderators, meta-moderators, and users. By decentralizing the moderation process to the user base, this approach has been successful in quickly and consistently distinguishing between high- and low-quality comments, reducing the burden on centralized staff to handle disruptive behaviors~\cite{lampe2004slash}. 

Slashdot's approach of granting power to standout community members encouraged contributions that meet the community's norms of quality~\cite{kraut2012building}. By distributing responsibility among its members, it encouraged collective action to improve the discussion space. Inspired by this, our goal is to adapt a version of this distributed model to commenting systems in order to build a collective discussion space that incorporates the complexity of viewpoints, shifting the focus from fragmented individual expressions to collaboratively structured and managed discussions. This approach empowers users to take on diverse roles in organizing the discussion space, preventing dominance by any single viewpoint and ensuring discussions reflect a broad range of perspectives. 

We propose a three-stage structure for collectively organizing discussions through distributed roles: \textbf{clustering}, \textbf{summarization}, and \textbf{threading}. Summaries have proven essential in previous research, synthesizing ideas with discussions and encouraging participants to reflect on the conversation~\cite{zhang2017wikum, nam2007arkose}, thus helping them actively engage with others' perspectives~\cite{kriplean2012you}. By integrating these summaries with clustering and threading features, we transform flat discussions into a multi-level structure, allowing users to easily navigate between perspectives and explore ideas in greater depth~\cite{hoque2015convisit, hoque2016multiconvis}. Clustering organizes related comments into groups, identifying recurring themes or shared viewpoints; summaries condense these clusters into concise insights; and threading organizes the clusters and summaries hierarchically. Through these features, the system collaboratively builds a collective understanding of complex viewpoints, creating a self-sustaining environment where every user contributes to the constructive exchange of opinions and maintains the quality of discussions through assigned activities, in line with a crowd-based moderation approach~\cite{diakopoulos2011towards}.

\subsection{Role Assignment}
To implement a distributed model, we introduce three distinct user roles (Level 0, 1, and 2) that collaborate on structuring and organizing of discussions, focusing on three main features~--~\textbf{clustering}, \textbf{summarizing}, and \textbf{threading}. These features are not assigned individually to each level but are collectively realized through the collaboration of different levels.
The role hierarchy in our system was designed so that users at each level take responsibility for increasingly broader units of the discussion. Level 0 users start by clustering individual comments. Level 1 users then review and summarize each cluster, working with ideas at a wider scope. Finally, Level 2 users create and organize threaded discussions based on those summaries. Assigning roles in this order ensures that each level’s participation is enabled by the validated outputs from the previous level and that collaboration emerges through cycles of proposing and reviewing across role boundaries.

\begin{figure*}
     \centering
     \includegraphics[width=0.7\textwidth]{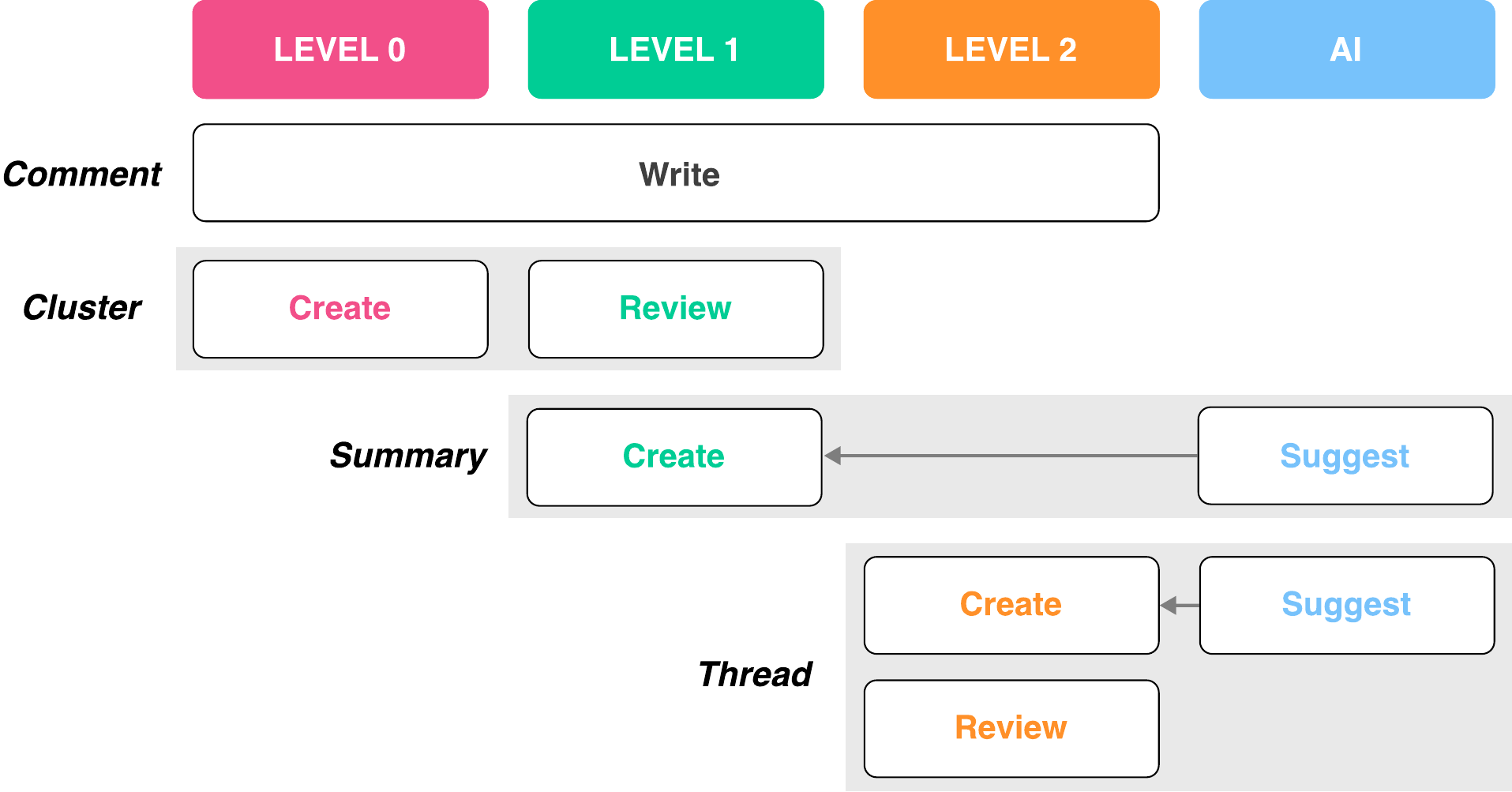}
     \caption{Assigned roles for each user level: Lower levels contribute to smaller units of the discussion space, ordered as clusters, summarizations, and threads. All users can comment, while clustering, summarizing, and threading are collaboratively created and reviewed across different user levels. AI assists by suggesting summarizations and thread topics.}
     \label{figure:role-assignment}
     \Description{
     The figure shows a diagram of user roles divided by level, including Level 0, Level 1, and Level 2, where each level has specific actions related to commenting, clustering, summarizing, and threading. Lower levels contribute to smaller units of the discussion space, ordered as clusters, summarizations, and threads. At the top, all levels are shown able to 'Write' in the comment section. Beneath that, to make a cluster, Level 0 users can 'Create' and Level 1 users can 'Review.' To create a summarization, Level 1 users can 'Create' with AI suggesting the summary in this process. To make a new thread, Level 2 users can 'Create' the thread topic and 'Review' the created topics, with AI also suggesting topics in this process.
     }
\end{figure*}

The assigned roles for each level are presented in Figure~\ref{figure:role-assignment}.

Lower levels are mapped to roles that contribute to smaller units of the discussion space. 
The smallest unit, the cluster, is managed by Level0~(LV0) users. LV0 users create clusters, which are then reviewed by Level1~(LV1) users. Once a cluster is accepted by a certain number of LV1 users, it is finalized and made visible to all participants. The task of summarizing these clusters is assigned to LV1 users. To assist in this task, the system uses AI suggestions to help LV1 users perform their summarization role more effectively, reducing the manual effort required for summarizing.

Level2~(LV2) users are responsible for creating new threads that introduce discussion topics and shape the high-level flow of the discussion. They can propose threads based on their perspectives, either branching off from existing discussions or addressing gaps in the current conversation, with AI suggestions assisting in the process. These proposed threads are then reviewed by other LV2 users, who decide whether to accept or decline them. The required number of approvals or denials for the creation of clusters and threads can be determined by considering the need for sufficient deliberation while avoiding significant delays in their creation. In this paper, we predefine this number as three, considering the scale of the experimental environment.

\subsection{Implementation}
We implemented our system through a browser extension. Since each news outlet has a different markup format, we chose CNN as the news outlet used for this study\footnote{https://edition.cnn.com/} and implemented the browser extension to work on top of its website. CNN was chosen both because of its familiarity to potential study participants and because it does not require a login or subscription for participants to access, unlike many other news outlets with paywalls. While the current implementation is tailored to CNN, the extension is designed to be adaptable and can be configured to work with most new outlet websites.
The backend of the system is developed using Python, with FastAPI as the web framework, Uvicorn as the web server, and SQLite as the database management system. The frontend is built using React.js, with Webpack utilized for module bundling and asset management.
In different parts of the system, we incorporated AI-generated content to serve as both an initial starting point and as suggestions to assist user activity. We used the \texttt{GPT-3.5-turbo model} to generate initial discussion topics, guiding questions, suggested summaries, and new topic suggestions. The prompts were iteratively refined by our research team through assessing the quality of the outputs. 
The final prompts used for each feature are detailed in Appendix~\ref{appendix:prompt}.

\subsection{System Features}
\begin{figure*}
     \centering
     \includegraphics[width=0.65\textwidth]{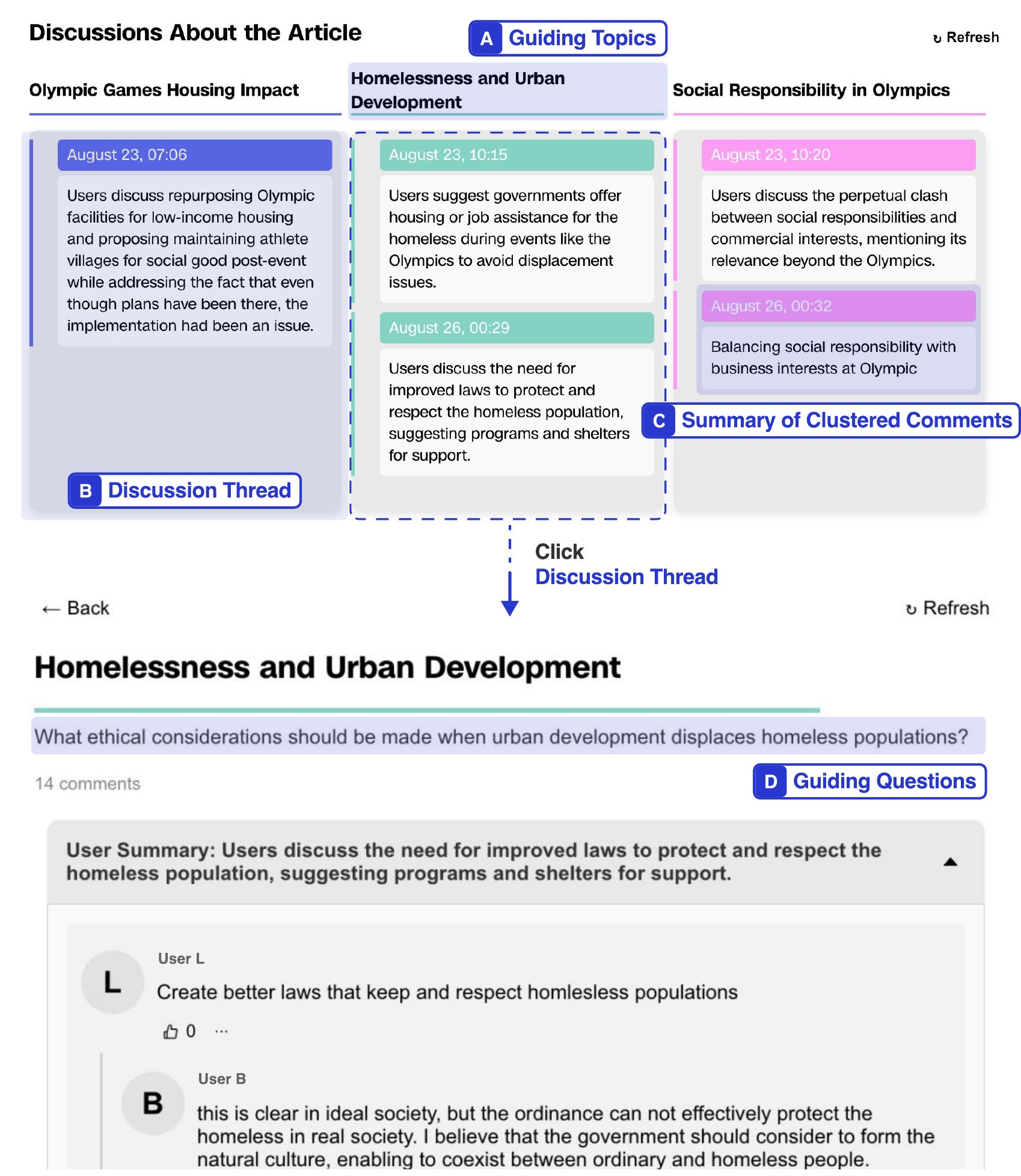}
     \caption{The overview of our system: (a) guiding discussion topics, three of which are initially generated by AI, (b) discussion thread showing the overview of the corresponding comment section for each discussion topic, (c) summaries of clusters displayed in each discussion thread, along with timestamps for the summaries created, and (d) guiding question to prompt the discussion in each discussion thread, generated when the topic is created}
     \label{figure:system-overview}
    \Description{
    The figure shows an interface for the system overview. At the top, three (a) guiding topics are displayed: 'Olympic Games Housing Impact', 'Homelessness and Urban Development', and 'Social Responsibility in Olympics', which are initially generated by AI. Beneath the guiding topics, there are corresponding (b) discussion threads, each showing the overview of a corresponding comment section. In these discussion threads, (c) summaries of clustered comments are shown along with the timestamps of their creation. In the figure, 'Balancing social responsibility with business interest at Olympic' is highlighted as an example of summary. After clicking on one of the discussion threads (e.g., 'Homelessness and Urban Development), the comment section appears, with (d) guiding questions at the top (e.g., 'What ethical considerations should be made when urban development displaces homeless populations?'), which are designed to prompt the discussion and are initially generated when the topics are created.
}
\end{figure*}

Upon enabling the extension, users can set their username and input their assigned user level. After clicking the button to add the discussion section, it is inserted into their browser's view of a CNN article page, as shown in Figure~\ref{figure:system-overview}. The first time the system is initialized on a given article's page, three AI-generated discussion topics, which we will refer to as \textbf{(a) guiding topics}, are generated. These topics are then shown identically to all subsequent users who join the discussion on the same article. Each discussion topic is given its own comment section, which we refer to as a \textbf{(b) discussion thread}. Each discussion thread includes \textbf{(c) summaries of clustered comments} created by users, if any, along with timestamps to indicate the sequence of the discussion flow. By clicking on each discussion thread, users can enter the comment section, where the \textbf{(d) guiding question} will appear at the top. This question, initially generated by AI at the time the discussion topic is created, is designed to help commenters begin the conversation during the initial phase of discussion. This guiding question is intended to serve only as a prompt, so the discussion is not required to remain confined to it. We have phrased this guiding question in neutral language, ensuring it remains open-ended and does not favor any particular viewpoint to avoid bias.

The comment section contains all basic commenting features including writing comments, replying to comments, liking other comments, editing, and deleting comments. According to their assigned roles, each user is presented with a system having different functionalities as outlined below to collectively achieve each feature of creating clusters, summarizations, and threads. 

\subsubsection{Clustering}
The \textit{clustering} feature is designed to merge comments that share similar themes or perspectives. In order to create clusters grouped by similar viewpoints, users are naturally encouraged to carefully read other comments and reconsider the issue from perspectives different from their own. This process not only promotes a broader perspective but also helps create coherent clusters of related comments, making it easier for all the users to follow the conversation and engage with content that aligns with their interests.

\begin{figure*}
     \centering
     \includegraphics[width=\textwidth]{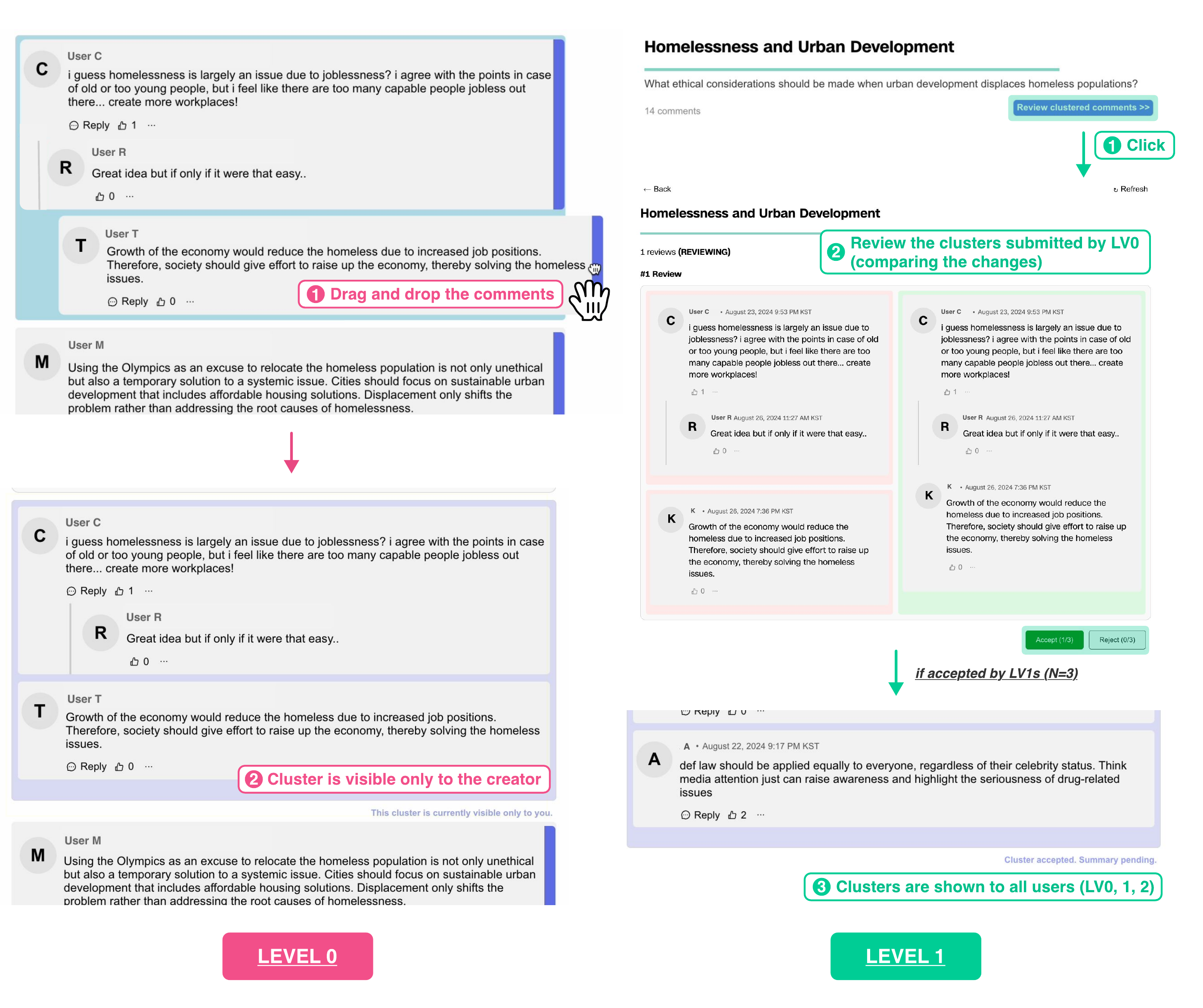}
     \caption{Workflow of Clustering: LV0 users propose clusters by dragging and dropping comments into the desired locations (left). LV1 users then review these clusters by comparing the changes before and after the clustering activity. The clusters become visible to all users once they are approved by the required number of LV1 users (right).}
     \label{figure:system-workflow-cluster}
     \Description{
     The figure shows the workflow of clustering in a system where Level 0 (LV0) and Level 1 (LV1) users collaborate together to create a cluster. The left side of the figure illustrates how LV0 proposes a cluster by dragging and dropping comments into desired locations. The interface displays individual comments, with LV0 users arranging them into clusters, which remain visible only to the creator during this stage.
     On the right side, after the clustering activity, LV1 users review the clusters by comparing the changes before and after clustering, with a side-by-side view of comments showing the 'Before' and 'After' state of clustering. LV1 users can either approve or decline each cluster, and once a certain number (N=3) of LV1 users agree, the clusters become visible to all users across levels (LV0, LV1, LV2). 
     The figure includes step-by-step instructions for this process: For LV0 users, (1) drag and drop the comments, and (2) the cluster remains visible only to the creator. For LV1 users, (1) click the 'Review Clustered Comments' button, (2) review the clusters by clicking 'Approve' or 'Decline' button, and (3) the clusters become visible to other levels after approved by the required number of LV1 users.
     }
\end{figure*}

Users assigned to LV0 and LV1 collaboratively create clusters, with LV0 proposing the clusters and LV1 reviewing the proposed clusters, as shown in Figure~\ref{figure:system-workflow-cluster}. Clustering can be done by dragging and dropping comments into the desired location through the system. Clusters can contain multiple comments; however, reply-level comments cannot be moved into a cluster separately. If a user clusters comments, the replies are moved along with them.

To review these clustering activities, LV1 users can access the review page by clicking the ``Review Clustered Comments'' button at the top right of their screen. A list of cluster reviews will be displayed, allowing LV1 users to compare the discussion space before and after clustering. The left side shows the comments before clustering, while the right side displays the updated arrangement. During the review process, LV1 users can approve or decline clustering activities. A cluster is displayed in the comment section once it has been approved by the required number of LV1 users and is removed if it is declined by the predefined number of users. In this paper, we set the requirement for both approval and denial to three participants, but the system can be adjusted to accommodate different thresholds based on their needs. 
The final clusters will be displayed in a blue box, visible to all users.

\subsubsection{Summarizing}
The \textit{summarizing} feature aims to distill clustered comments into a single, cohesive summary that encapsulates the core ideas of the discussion. This feature is intended to minimize redundancy and ensure that key points are highlighted, preventing it from being overshadowed by repetitive comments. By streamlining the conversation, this feature enables participants to quickly identify emerging trends and common concerns of discussed issues, making it easier to engage with the core aspects being presented.

\begin{figure*}
     \centering
     \includegraphics[width=\textwidth]{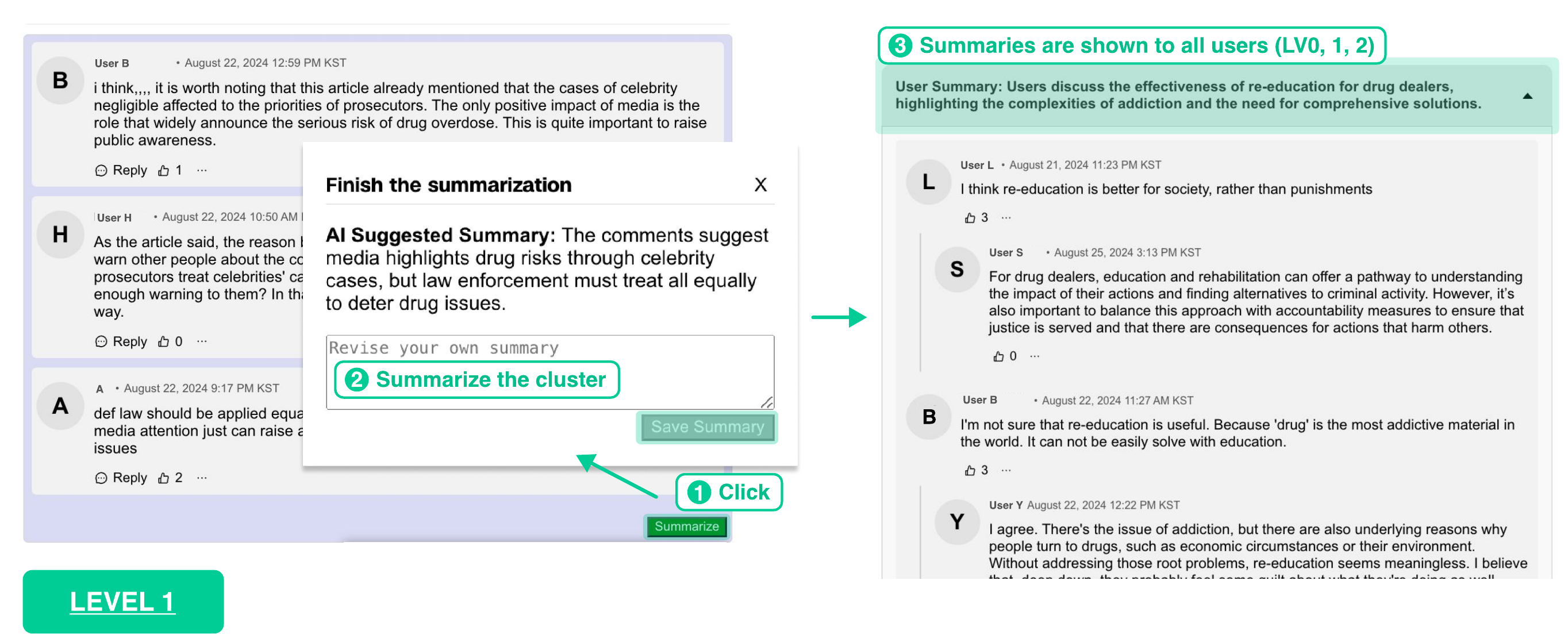}
     \caption{Workflow of Summarizing:~(1)~LV1 users summarize accepted clusters by clicking the `Summarize' button within the cluster.~(2)~A modal will display an AI-generated summary suggestion, which users can revise or replace with their own (left).~(3)~Once finalized, the summary becomes visible to all users and is displayed at the top of the cluster (right).}
     \label{figure:system-workflow-summary}
     \Description{
     The figure shows the workflow for summarizing accepted clusters in a system by Level 1 (LV1) users. On the left, there is a `Summarize' button within a cluster for LV1 users. After clicking it, a modal appears displaying an AI-generated summary suggestion. The suggestion says: 'The comments suggest media highlights drug risks through celebrity cases, but law enforcement must treat all equally to deter drug issues.' Users can revise or replace the AI-generated summary with their own text by typing in the provided field. On the right side, the final summary is displayed at the top of the clustered comments once it is saved. Below the summary, users' original comments that were included in the clusters are visible.
     The figure includes step-by-step instructions for this process: (1) LV1 users click the 'Summarize' button, (2) a modal with the AI-suggested summary appears, allowing the user to edit or replace it with their own summarization, and (3) the finalized summary becomes visible to all users and is displayed at the top of the cluster.
}
\end{figure*}

After clusters are created per the process described above, LV1 users can summarize accepted clusters. Upon clicking the ``Summarize’’ button in the cluster, a pop-up modal appears with an AI-suggested summary (Figure~\ref{figure:system-workflow-summary}). While users can use the suggested summary, they are encouraged to revise or create their own to critically engage with the discussion, ensuring that the final summary reflects a comprehensive range of viewpoints. Once an LV1 user has created a summary, it will appear at the top of the cluster and be visible to all users. After a cluster has been summarized, no additional comments can be added, as this would affect the appropriateness of the existing summary. Thus, when creating summaries, LV1 users are instructed to check that all relevant perspectives are included in the cluster, and assess if there is room for new comments. The summary of the cluster is displayed under the discussion thread on the first page, thereby informing users of the key discussions emerging in each thread.

\subsubsection{Threading}
The \textit{threading} feature organizes the comments into distinct threads based on specific aspects of the discussion, helping users focus on major perspectives and consider the issue in greater detail. This feature allows participants to easily navigate and contribute to specific lines of thought, encouraging more focused dialogue. Initially, three guiding topics are provided for the discussion section, with users having the flexibility to add additional topics.

\begin{figure*}
     \centering
     \includegraphics[width=\textwidth]{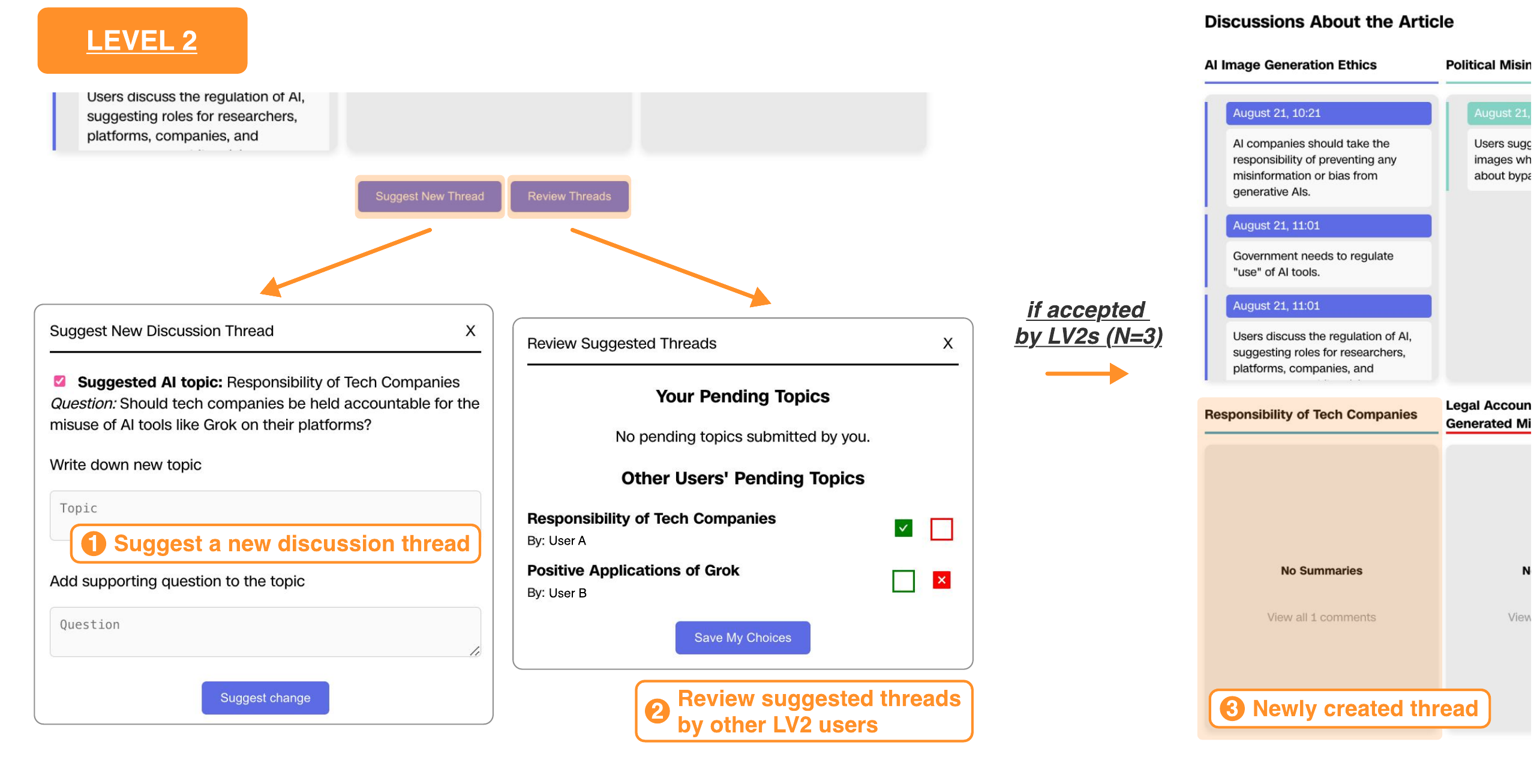}
     \caption{Workflow of Threading:~(1)~LV2 users propose new thread topics by selecting from AI-suggested topics or by suggesting their own.~(2)~Users review these topics by approving or declining each proposed thread.~(3)~A new discussion thread is created and becomes visible to all users once a topic is approved by the required number of LV2 users.}
     \label{figure:system-workflow-thread}
     \Description{
     The figure shows the workflow for threading in a system where Level 2 (LV2) users propose and review new discussion threads. On the left side, LV2 users are presented with the option to 'Suggest New Thread' by either selecting an AI-suggested topic or writing their own. The AI-suggested topic shown is 'Responsibility of Tech Companies', with the supporting question: 'Should tech companies be held accountable for the misuse of AI tools like Grok on their platforms?' Users can write their own new topic and add a supporting question in the provided fields.
     In the middle of the figure, a 'Review Suggested Threads' section allows LV2 users to approve or decline pending topics proposed by others. Two topics are shown: 'Responsibility of Tech Companies' and 'Positive Applications of Grok', with checkmark and cross icons for approval or rejection.
     On the right side, once a thread is accepted by the required number of LV2 users (N=3), the new discussion thread appears under the 'Discussions About the Article' section. The newly created thread, titled 'Responsibility of Tech Companies', is now visible to all users.
     The figure includes step-by-step instructions for this process: (1) LV2 users suggest new thread topics, (2) review suggested topics by other LV2 users, and (3) a new thread is created and becomes visible once it is approved by the required number of LV2 users.
}
\end{figure*}

New discussion thread topics are created through the suggestions and review activities of LV2 users, as outlined in Figure~\ref{figure:system-workflow-thread}. As discussions within each topic evolve, LV2 users can propose new topics for threads by clicking the ``Suggest New Thread’’ button. A pop-up modal will appear with an AI-generated topic suggestion and a guiding question for the new discussion, designed to assist users in formulating new topics. These topics and guiding questions are generated based on the article's content. Users have the option to select the AI-suggested topic by checking the box, but they are encouraged to suggest a topic that incorporates and aligns with the ongoing discussion flow. 

To review these created threads, users click the ``Review Threads'' button. A pop-up modal will display a list of suggested topics from other users, which can be approved or declined via checkboxes. A new discussion thread is created once a topic is approved by the required number of LV2 users. We set this requirement to three participants for the purpose of the user study, but it can be adjusted to different thresholds. The newly accepted threads will appear at the bottom of the original thread boxes on the first page. All users can access these newly created threads and participate in the discussion.

\section{Method}

\subsection{Participants}
We recruited a total of 40 participants to use the commenting system browser extension through communities managed by our institution. We asked them to fill out a survey asking about the frequency of reading news articles and writing comments and their motivations to do those activities. We filtered out the participants who never read articles during the week to verify our system among those who regularly consume news articles. 
All study tasks, including reading news articles and writing comments or summaries, were conducted in English. For recruitment, we specified that participation required the ability to understand English news content and produce written responses in English, and participants confirmed that they were able to do so as part of the screening survey. Participants used English for all written interactions within the system.
Two participants dropped out due to scheduling constraints, so a total of 38 participants (Age=19-31, $M=23.12$, $SD=3.25$;
22 male and 16 female) were included in our study. Participants received 40,000KRW for their participation over six days, with an expected usage of 10 to 20 minutes of the system each day.

\subsection{Study Procedure}
\subsubsection{System Usage}
For onboarding, we provided participants with an instructional document that they could reference throughout the study period. Additionally, we provided a short video explaining how to use the system features specific to their assigned roles before beginning the user study.

Each participant was assigned a specific level prior to the start of the user study, and this level remained constant throughout the experiment. We conducted a within-subject study comparing our system with a baseline. For the baseline, we implemented a system that displays the standard commenting section, featuring only the commenting, reply, edit, delete, and like functions that users are familiar with. The order in which participants used the baseline system and our system was counterbalanced.

The study was conducted using a total of six articles from CNN, with two articles assigned to each of three different topics: \textit{Technology}, \textit{Crime}, and \textit{Economy}. The articles were carefully selected by the research team based on their societal impact and the level of public interest they were likely to generate. The selection prioritized articles likely to provoke differing viewpoints and those that provide enough detail to highlight the diversity of perspectives across different social groups, aiming to observe how collective discourse is shaped through the system. To minimize potential learning effects during the user study when testing two articles on the same topic with different systems, we chose articles that highlight different aspects while covering the same issue. These different aspects were not defined by opposing stances toward the issue.
For example, within the first topic covering economic issues related to hosting the Olympics, the first article focuses on the challenges of relocating the homeless and the issue of gentrification, while the second article addresses the financial costs of hosting the Olympics and the sustainability goals associated with the event. The specific topics covered by each pair of articles, along with the titles of the articles used in the study, are listed below:

\begin{enumerate}[itemsep=0.5ex]
    \par\medskip
    \item \textbf{[Technology] Concerns Over Emerging AI Technologies and Their Impact}
    \par\smallskip
    \begin{itemize}
        \item (System) \textit{``Elon Musk’s AI photo tool is generating realistic, fake images of Trump, Harris, and Biden''}
        \item (Baseline) \textit{``OpenAI worries people may become emotionally reliant on its new ChatGPT voice mode''}
    \end{itemize}
    \par\medskip
    \item \textbf{[Crime] Legal and Ethical Issues Surrounding Drug-Related Deaths and Treatments}
    \par\smallskip
    \begin{itemize}
        \item (System) \textit{``Why it’s important to prosecute celebrity drug deaths and the message it sends, according to legal experts''}
        \item (Baseline) \textit{``Even before Matthew Perry’s death, experts worried about the ‘Wild West’ of ketamine treatment''}
    \end{itemize}
    \par\medskip
    \item \textbf{[Economy] Financial Challenges of Hosting the Olympics}
    \par\smallskip
    \begin{itemize}
        \item (System) \textit{``Paris continues a shameful Olympic tradition''}
        \item (Baseline) \textit{``Hosting the Olympics has become financially untenable, economists say''}
    \end{itemize}
    \par\medskip
\end{enumerate}

We did not randomize the order of the articles, as we prioritized evenly spacing articles on the same topic to maintain consistent intervals across all three topics, which we expected to have a greater impact than the order of topics.
Each article started with no comments present when viewed by the first assigned group, but all comments and interactions from the first group remained visible for the second group. This approach is designed to observe the progression of discussions as participant engagement increases over time, from the initial publication stage of the article to its expanded discourse phase.

\begin{figure*}
     \centering
     \includegraphics[width=0.8\textwidth]{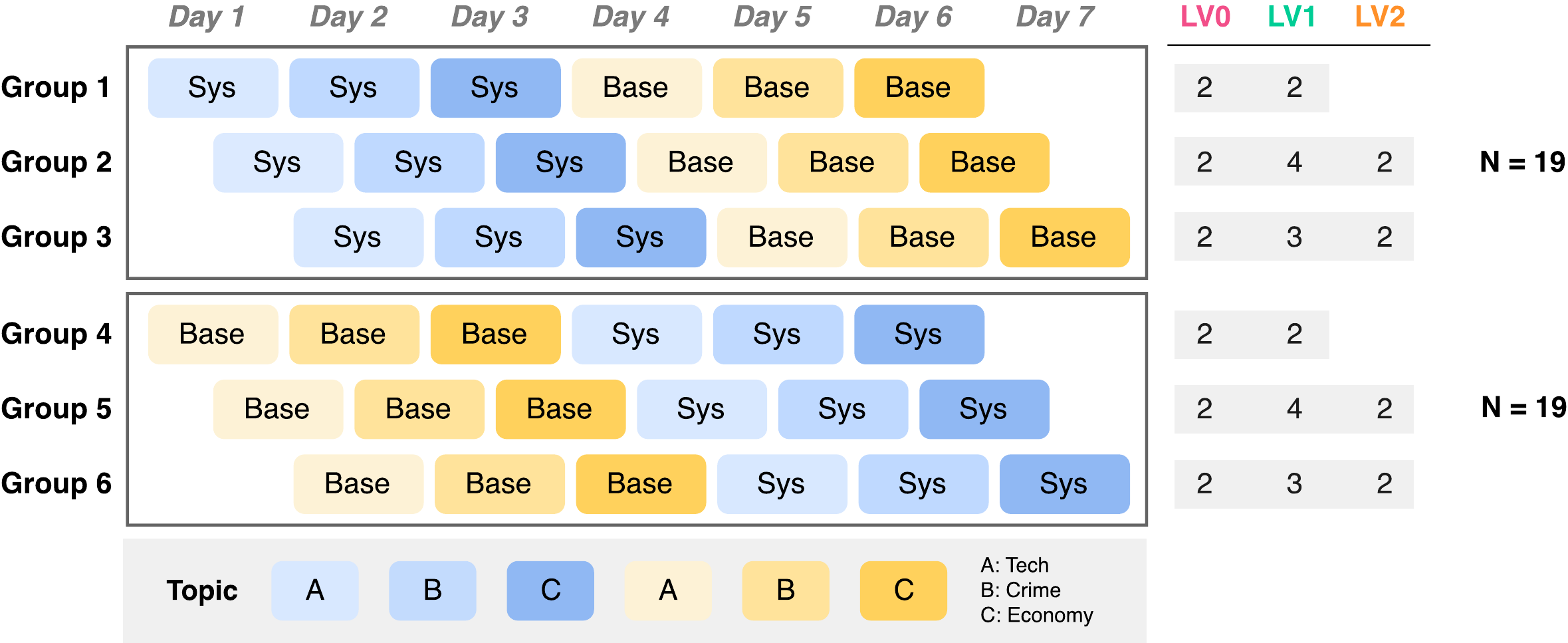}
     \caption{Procedure of the study. We conducted a within-subject study over six days for each participant, using both our system and a traditional commenting system, with the order balanced. The study included six articles, two for each of three different topics, with the same topics spaced evenly. Participants were divided into six groups with staggered start times to have all levels function simultaneously. The number of levels assigned to each group is shown on the right.}
     \label{figure:study-procedure}
     \Description{The figure displays the procedure of a within-subject study conducted over seven days, involving six participant groups. Each group alternates between using the system ('System' in blue) and a traditional commenting system ('Baseline' in yellow) across six days. The columns represent days, and each row represents a group. Groups 1 to 3 begin with the system, while groups 4 to 6 begin with the traditional commenting system, with alternating use on different days. The system and baseline are evenly distributed across days to ensure balance.
     On the right side of the figure, the number of users assigned to each level (LV0, LV1, LV2) is shown for each group. Group 1 include 2 LV0, 2 LV1 users, group 2 include 2 LV0, 4 LV1, 2 LV2 users, and group 3 include 2 LV0, 3 LV1, 2 LV2 users. Similarly, groups 4 to 6 also show the same distribution as groups 1 to 3, with a total of 38 participants participating in the study.
     At the bottom, a color-coded key shows the three topics discussed during the study: A (Tech), B (Crime), and C (Economy), with these topics spaced evenly throughout the study. The six groups have staggered start times to make all levels function simultaneously. }
\end{figure*}

Participants were divided into a total of six groups, balancing the number of participants at each level. Each group started at a different time, with a 12-hour interval between the start times. Since each level of participants has different needs to facilitate discussion in the system, the goal of assigning groups with different participation time periods was to have all levels work simultaneously. 
The number of participants at each level was adjusted based on the participation levels and ongoing activities observed during the pilot study. Figure~\ref{figure:study-procedure} summarizes the study procedure including group assignments, level distribution, article sequence and staggered participation schedule. After using the system, we asked participants to complete a post-survey about their general experiences and the impact of using our system. The post-survey questions are presented in Appendix~\ref{appendix:post-survey}.

\subsubsection{Interview}
We conducted follow-up interviews with participants who expressed an interest in participating in the interviews. These interviews were conducted via Zoom video calls and lasted between 20 and 40 minutes. A total of 14 participants took part. 
The distribution of assigned levels among the interview participants was as follows: 4 at LV0, 7 at LV1, and 3 at LV2. Participants were compensated with an additional 10,000KRW for their involvement in the interviews.

The interview was conducted in two phases. First, participants were asked to provide a brief overview of their experience using both systems, serving as a reminder of their overall experience. Next, we asked detailed questions about the impact of our system on the comment writing experience, based on the results of the post-survey. The overall structure of the interview questions is presented in the Appendix~\ref{appendix:interview}. Additional questions were asked based on the participants' responses. The interviews were primarily conducted in English; however, participants had the option to conduct the interview in Korean if they were not fluent in using English. The interview data were transcribed and translated into English for the analysis. We conducted a thematic analysis using an inductive approach, developing and refining codes for each category of questions~\cite{elo2008qualitative}. Initial coding generated 4–6 themes per category, and after multiple rounds of iteration, we combined them into three final themes for each category.

\subsection{Measures and Hypotheses}\label{section:4.3}
To evaluate the effects of our system relative to the baseline, we examined both engagement metrics and the quality of user comments. Engagement was measured through \textbf{comment length} (average word count) and \textbf{endorsement} (average likes received per comment), as these metrics are commonly used to capture patterns of participation and peer recognition in online discussions \cite{tenenboim2022comments}. To assess comment quality, we focused on four complementary dimensions that have been emphasized in prior work on deliberative and news comment spaces: 

\begin{itemize}
    \item \textbf{Perspective Diversity}: the extent to which discussions contained a balanced representation of viewpoints, measured using entropy- and coverage-based metrics of semantic clusters. Prior work has highlighted perspective diversity as a key outcome of online deliberation and comment moderation systems \cite{munson2010presenting}.
    \item \textbf{Argument Strength}: the proportion of claims accompanied by explicit reasoning or evidence, assessed through supported-claim ratios. This measure captures argumentative robustness, which has been central to evaluating deliberative quality in online spaces \cite{stromer2007measuring}.
    \item \textbf{Emotional Expression}: the prevalence of affective content, including overall emotionality as well as positive and negative emotions, derived from sentence-level emotion classification. Emotional tone is widely used to understand online discourse quality, with prior work showing how emotion shapes participation and civility \cite{chmiel2011negative}.
    \item \textbf{Politeness}: the use of prosocial language strategies, measured through politeness scoring. Politeness has been shown to contribute to sustained and constructive participation in online communities \cite{danescu2013computational}.
\end{itemize}

We derived the following hypotheses by linking these measures to the design intentions of our system. 
First, our system reduces the need for participants to individually reconstruct the shared context of the discussion. By presenting a pre-structured environment, where related ideas are already grouped and prior contributions are visible in an organized manner, the system eliminates redundant cognitive work typically required in free-form commenting. Guided prompts further streamline expression by helping participants focus on contributing new information or clarifying ideas. As a result, we expect that users can participate more frequently yet more concisely, without compromising the clarity or perceived value of their comments.

\begin{description}
    \item[H1.] Comments written with our system will be shorter than those in the baseline condition, while receiving a similar number of likes.
\end{description}

Second, clustering and summarization improve the overall visibility of distinct viewpoints. By organizing contributions into a clear structure, the system makes the range of perspectives easier to perceive and navigate, which we expect will support broader engagement with diverse ideas.

\begin{description}
    \item[H2.] Discussions in the system condition will exhibit greater perspective diversity than in the baseline condition. 
\end{description}

Third, guided prompts and distributed roles support users in constructing clearer and more analytically grounded arguments. By directing attention toward justification and evidence, the system encourages reasoning that is more explicit and well-supported, therefore improving argument strength.

\begin{description}
    \item[H3.] Comments in the system condition will demonstrate higher argument strength, with a greater proportion of claims supported by evidence.
\end{description}

Fourth, the system guides participation toward task-oriented interactions, which are expected to reduce opportunities for unstructured, reactive exchanges that often escalate negativity. By providing clear contribution pathways and shared conversational targets, users can focus on collaborative meaning-making rather than interpersonal conflict. We therefore expect emotional expression to remain constructive, with increased politeness in user interactions.

\begin{description}
    \item[H4.] Comments in the system condition will express less negative emotion and maintain or increase positive emotional expression compared to baseline.
    \item[H5.] Comments in the system condition will exhibit higher levels of politeness than those in the baseline condition.
\end{description}

Together, these hypotheses capture our expectation that the system would foster more balanced, civil, and analytically grounded discussion.

\section{Results}

In this section, we present the findings from the user study and follow-up interviews, organized by research questions.
We begin by presenting quantitative results comparing user activity and comment quality between the baseline and our system.
Next, we describe findings from the interviews, where participants described how the system supported different stages of the commenting process, and we present common themes of its contributions and limitations.

\subsection{RQ1. How does a structured discussion system influence patterns of user engagement?}

We first present an analysis of quantitative measures comparing the baseline and our system, including engagement metrics and statistics on the activities conducted by users at each level within our system.

\subsubsection{User Engagement Metrics}

Table~\ref{table:engagement-metrics} summarizes the engagement metrics comparing our system and baseline conditions across the three topics. Participants authored more comments in the system condition overall (a total of 230 in the system condition and 189 in the baseline), with increases in Technology and Crime and comparable counts in Economy. Reply counts were broadly similar between conditions, and the average number of likes per comment was lower in the system across topics. Comments were consistently shorter in the system condition, with lower average word counts for all three topics.

Since participants contributed comments in both conditions, we analyzed likes and comment length using multilevel models to account for the non-independence of observations(Table~\ref{table:nbglmm_effects}). Specifically, we fit generalized linear mixed models (NB-GLMMs) with condition (baseline and system) as a fixed effect and random intercepts for participants. We adopted this model since the diagnostic checks indicated overdispersion, with the variance of the data exceeding the mean. Comments in the system condition received fewer likes on average (IRR = 0.82, 95\% HDI [0.56, 1.15]), but this difference was not significant. For word count, system comments were significantly shorter (IRR = 0.74, 95\% HDI [0.66, 0.82]), corresponding to an approximate 26\% reduction relative to baseline. 
These findings suggest that while the system encouraged more frequent and concise contributions, it did not significantly alter patterns of peer endorsement.
This result is consistent with H1, indicating that the structured environment increased active but concise participation without reducing the perceived value of comments.

\begin{table*}[t]
\centering
\small
\begin{tabular}{lcccccc}
\toprule
 & \multicolumn{2}{c}{Technology} & \multicolumn{2}{c}{Crime} & \multicolumn{2}{c}{Economy} \\
\cmidrule(lr){2-3} \cmidrule(lr){4-5} \cmidrule(lr){6-7}
\textbf{Metric} & \textbf{Baseline} & \textbf{System} & \textbf{Baseline} & \textbf{System} & \textbf{Baseline} & \textbf{System} \\
\midrule
Total comments & 61 & 89 & 67 & 79 & 61 & 62 \\
Total replies & 35 & 36 & 40 & 38 & 33 & 25 \\
Average like count (SD) & 0.75 (1.13) & 0.53 (1.08) & 0.79 (1.52) & 0.65 (1.04) & 0.77 (1.37) & 0.58 (1.56) \\
Average word count (SD) & 47.52 (24.70) & 34.02 (17.22) & 49.36 (32.47) & 33.14 (17.48) & 50.51 (29.78) & 37.68 (25.29) \\
\bottomrule
\Description{This table compares engagement metrics between the system and baseline across three article topics. It shows an increase in commenting activity for all topics, with 61, 67, and 61 comments for the baseline, and 89, 79, and 62 comments for the system (in order of tech, crime, and economy). The levels of replies and likes remained similar across all topics. Additionally, the average comment length and word count were significantly shorter for all three topics, with an average word count of 49.13 for the baseline and 34.96 for the system.}
\end{tabular}
\caption{Engagement metrics by topic and condition. Counts are totals; likes and word count are means (SD).}
\label{table:engagement-metrics}
\end{table*}

\begin{table*}[t]
\centering
\small
\begin{tabular}{llllcc}
\toprule
\textbf{} & \textbf{Model} & \textbf{IRR} & \textbf{95\% HDI} \\
\midrule
Like count & NB-GLMM & 0.817 & [0.561, 1.152] \\
Word count & NB-GLMM & 0.736 & [0.656, 0.824] \\
\bottomrule
\Description{This table compares comments made in the system condition with those in the baseline condition. It reports two measures: the number of likes and the length of comments. On average, comments in the system condition received about 18\% fewer likes than baseline comments, but this difference was not statistically meaningful. In contrast, system comments were about 26\% shorter than baseline comments, and this reduction was statistically meaningful.}
\end{tabular}
\caption{Results of negative binomial generalized linear mixed models (NB-GLMMs) comparing system and baseline conditions. 
IRRs (Incidence Rate Ratios) less than 1 indicate reductions in the system condition relative to baseline. 
System comments were significantly shorter than baseline comments, while differences in like counts were not significant.}
\label{table:nbglmm_effects}
\end{table*}

\subsubsection{Details of System Usage}
This section presents a detailed analysis of how users interacted with the system, highlighting their activities and the resulting outcomes across the three core functionalities: clustering, summarization, and thread creation. The Table~\ref{table:system-summary} presents a detailed summary of these three activities and their outcomes across different articles. The findings illustrate how the balanced tension among activities performed by users at different levels contributes to an active and deliberate process shaping the discussion space.

\textbf{Clustering} Each article demonstrated a similar trend, with 5 to 7 finalized clusters emerging as the outcome of extensive clustering activities. Notably, the formation of these completed clusters required 16 to 36 total individual clustering activities per article, highlighting the intensive review and refinement process involved. Pending clusters, which had not yet reached final approval, predominantly exhibited an Accept(2/3) status, suggesting that many were close to completion. For example, in the ``Tech'' article, 10 of the 12 pending clusters were nearing acceptance, while the ``Crime'' article showed 8 out of 12 pending clusters in the same state. The ``Economy'' article displayed a smaller scale of activity, with both of its 2 pending clusters marked as Accept(2/3). The average of 9.67 activities resulted in accepted clusters, while 9.33 activities concluded with denial, indicating significant deliberation and evaluation among LV1 users during the clustering process. These findings highlight that, although the finalized clusters represent a smaller subset of the total activities, the dynamics between suggestions from LV0 contributors and evaluations by LV1 reviewers created a balanced and constructive tension.

\begin{table*}[]
\centering
\resizebox{\textwidth}{!}{%
\begin{tabular}{l|ccccc|c|cccc}
\toprule
 & \multicolumn{5}{c|}{\textbf{Clustering}} & \multicolumn{1}{c|}{\textbf{Summary}} & \multicolumn{4}{c}{\textbf{Threading}} \\ \midrule
 & \multicolumn{1}{l|}{\cellcolor[HTML]{C0C0C0}\begin{tabular}[c]{@{}l@{}}Total\\ number of\\ created\\ clusters\end{tabular}} &
   \begin{tabular}[c]{@{}l@{}}Total\\ number of\\ clustering\\ activities\end{tabular} &
   \begin{tabular}[c]{@{}l@{}}Accepted\\ clustering\\ activities\end{tabular} &
   \begin{tabular}[c]{@{}l@{}}Pending\\ clustering\\ activities\end{tabular} &
   \begin{tabular}[c]{@{}l@{}}Denied\\ clustering\\ activities\end{tabular} &
   \cellcolor[HTML]{C0C0C0}\begin{tabular}[c]{@{}l@{}}Total\\ number of\\ created\\ summaries\end{tabular} &
   \multicolumn{1}{l|}{\cellcolor[HTML]{C0C0C0}\begin{tabular}[c]{@{}l@{}}Total\\ number of\\ suggested\\ threads\end{tabular}} &
   \begin{tabular}[c]{@{}l@{}}Accepted\\ threads\end{tabular} &
   \begin{tabular}[c]{@{}l@{}}Pending\\ threads\end{tabular} &
   \begin{tabular}[c]{@{}l@{}}Denied\\ threads\end{tabular} \\ \midrule
\textbf{Tech} & \cellcolor[HTML]{C0C0C0} 6 & 36 & 11 & 12 & 13 & \cellcolor[HTML]{C0C0C0} 6 & \cellcolor[HTML]{C0C0C0} 4 & 3 & 1 & 0 \\ \midrule
\textbf{Crime} & \cellcolor[HTML]{C0C0C0} 7 & 31 & 11 & 12 & 8 & \cellcolor[HTML]{C0C0C0} 2 & \cellcolor[HTML]{C0C0C0} 4 & 3 & 1 & 0 \\ \midrule
\textbf{Economy} & \cellcolor[HTML]{C0C0C0} 5 & 16 & 7 & 2 & 7 & \cellcolor[HTML]{C0C0C0} 5 & \cellcolor[HTML]{C0C0C0} 3 & 1 & 2 & 0 \\ \bottomrule
\end{tabular}%
}
\vspace{8pt}
\caption{Summary of system usage statistics, including the number of clusters, summaries, and threads created by users. A breakdown of activities for creating clusters and threads is provided, detailing the number of review activities conducted for these creations. The total number of activities is presented, divided into accepted, pending, and denied for both clusters and threads.}
\label{table:system-summary}
\Description{This table presents a summary of our system usage statistics, separated by clusters, summaries, and threads. It shows the number of clusters, summaries, and threads created by users across all three articles. For clustering and threading activities, a breakdown is provided, detailing the number of review activities conducted for these creations. The total number of activities is divided into accepted, pending, and denied for both clusters and threads.
On average, 6 clusters, 4.3 summaries, and 2.3 threads were created per article. For clustering activities, an average of 27.7 clustering activities were conducted to create the total number of clusters, having similar levels of accepted and denied clustering activities.}
\end{table*}

\textbf{Summarization} Across the three articles, a total of 6, 2, and 5 summaries were created for ``Tech'', ``Crime'', and ``Economy'', respectively. In the ``Tech'' and ``Economy'' articles, all finalized clusters were converted into summaries. In contrast, the ``Crime'' article showed only 2 out of 7 clusters resulting in summaries. This indicates that the clusters undergo detailed assessment to ensure that all relevant perspectives are included before being converted into summaries, reflecting a deliberate review process by LV1 reviewers. The timing of summary creation shows that most summaries (10 out of 13) were generated within 48 hours of initial activity on the article. This suggests that the first 48 hours represent a critical window for generating diverse and non-redundant discussions, resulting in well-rounded summaries that effectively capture the essence of the clustered content.

\textbf{Threads} Across the three articles, a total of 3, 3, and 1 threads were created through user-suggested topics proposed by LV2 users. In both the "Tech" and "Crime" articles, one of the created threads included an AI-suggested topic. Pending topics still under review include 1, 1, and 2 threads for ``Tech'', ``Crime'', and ``Economy'', respectively, with the pending topic in ``Crime'' also containing an AI-suggested topic. The use of AI-suggested topics across all three articles highlights their relevance with the ongoing discussion flow as perceived by users. However, the observation that AI-suggested topics were not always the first to be created suggests that users remain actively engaged in introducing their own perspectives and initiating discussions. We present the details of the threads including initial topics provided by the system, user-generated topics for newly created threads, and the pending topics under review in Table~\ref{table:threadtopics}.

\newcolumntype{Y}{>{\raggedright\arraybackslash}X}

\begin{table*}[h!]
\centering
\small
\begin{tabularx}{\textwidth}{lYYY}
\toprule
 & \textbf{Initial topics} & \textbf{Created thread topics} & \textbf{Pending topics} \\
\midrule
\textbf{Tech} &
AI Image Generation Ethics; Political Misinformation Online; Impact of AI on Elections &
Responsibility of Tech Companies \textit{(AI-suggested)}; Legal Accountability for AI-Generated Misinformation; Misuse of AI-based Images &
Positive Applications of Grok \\
\addlinespace[2pt]
\hline
\addlinespace[2pt]
\textbf{Crime} &
Celebrity Drug-Related Deaths; Accountability of Drug Dealers; Publicity and Prosecution &
Legal Proceedings for Drug-Related Deaths; Reducing Overdose Incidents; The way for reducing the accident of overdose,
Medical professional responsibility \textit{(AI-suggested)} &
Medical Policy \\
\addlinespace[2pt]
\hline
\addlinespace[2pt]
\textbf{Economy} &
Olympic Games Housing Impact; Homelessness and Urban Development; Social Responsibility in Olympics &
Other Issues Regarding the Olympics &
Gentrification Effects of the Olympics \textit{(AI-suggested)}; Building infrastructure only once to have a permanent location to host all types of sport \\
\bottomrule
\end{tabularx}
\vspace{8pt}
\caption{Thread topics for the three articles, shown left to right as: (1) initial topics provided by the system, (2) user-generated topics accepted by other users, and (3) pending topics under review. AI-suggested topics are indicated in italics.}
\label{table:threadtopics}
\Description{This table lists thread topics for Technology, Crime, and Economy articles. For each article, columns present the system-provided initial topics, user-created topics that were accepted, and topics still pending; AI-suggested items are marked in italics.}
\end{table*}

\subsection{RQ2. In what ways does the system affect the quality of user comments?}
To better understand how the system influenced the qualities of user comments, we assessed four complementary measures introduced in Section~\ref{section:4.3}: \textbf{Perspective Diversity}, \textbf{Argument Strength}, \textbf{Emotional Expression}, and \textbf{Politeness}. 

\begin{table*}[t]
\centering
\label{tab:results}
\begin{tabular}{lcccccc}
\toprule
\textbf{Measure} & \textbf{Baseline} & \textbf{System} & \boldmath$\Delta$ & \boldmath$\beta$ (SE) & 95\% CI & Sig. \\
\midrule
\multicolumn{7}{l}{\textit{Perspective Diversity}} \\
\hspace{1em}H\_norm & 0.413 & 0.483 & +0.070 & 0.070 (0.030) & [0.012, 0.129] & \textbf{p = .018} \\
\hspace{1em}Coverage & 0.433 & 0.466 & +0.033 & 0.033 (0.033) & [$-$0.032, 0.099] & n.s. \\
\midrule
\multicolumn{7}{l}{\textit{Argument Strength}} \\
\hspace{1em}SCR & 0.228 & 0.153 & $-$0.075 & $-$0.075 (0.027) & [$-$0.128, $-$0.021] & \textbf{p < .01} \\
\midrule
\multicolumn{7}{l}{\textit{Emotion}} \\
\hspace{1em}Overall Emotionality & 0.656 & 0.552 & $-$0.104 & $-$0.109 (0.019) & [$-$0.147, $-$0.071] & \textbf{p < .001} \\
\hspace{1em}Joy & 0.009 & 0.003 & $-$0.006 & $-$0.529 (0.092) & [$-$0.709, $-$0.350] & \textbf{p < .001} \\
\hspace{1em}Sadness & 0.033 & 0.008 & $-$0.024 & $-$0.590 (0.150) & [$-$0.884, $-$0.296] & \textbf{p < .001} \\
\hspace{1em}Fear & 0.009 & 0.005 & $-$0.004 & $-$0.512 (0.117) & [$-$0.741, $-$0.282] & \textbf{p < .001} \\
\hspace{1em}Surprise & 0.010 & 0.006 & $-$0.004 & $-$0.383 (0.113) & [$-$0.605, $-$0.162] & \textbf{p < .01} \\
\hspace{1em}Anger & 0.003 & 0.004 & 0.001 & $-$0.069 (0.093) & [$-$0.251, 0.112] & n.s \\
\hspace{1em}Disgust & 0.002 & 0.005 & 0.002 & $-$0.009 (0.081) & [$-$0.167, 0.149] & n.s \\

\midrule
\multicolumn{7}{l}{\textit{Politeness}} \\
\hspace{1em}Score & 0.573 & 0.581 & +0.007 & 0.007 (0.009) & [$-$0.011, 0.025] & n.s. \\
\bottomrule
\end{tabular}
\vspace{8pt}
\caption{Summary of key outcome measures across conditions. Baseline mean reflects the intercept; $\Delta$ indicates the change under the system condition. Significant effects are bolded.}
\Description{The table shows the summary of the key outcome measures across the conditions, along the four measures of perspective diversity, argument strength, emotion, and politeness. The baseline mean reflects the intercept and the change under the system condition is denoted as delta. We observed significant results in all measures except for emotions of anger and disgust, and politeness.}
\end{table*}

\subsubsection{Perspective Diversity}
To assess whether the system fostered greater diversity of perspectives within discussions, we embedded all comments using all-MiniLM-L6-v2 (Sentence-BERT) and clustered them into semantic groups using K-means. For each article–condition pair, we computed two standard entropy-based measures: normalized entropy (H\_norm), capturing the evenness of distribution across clusters, and coverage, reflecting the proportion of clusters represented. Mixed-effects linear models were fit with condition as a fixed effect and article as a random intercept. Results showed that H\_norm was significantly higher in the system condition ($\beta$ = +0.070, 95\% CI [0.012, 0.129], z = 2.38, p = .018), indicating more balanced participation across perspectives. By contrast, coverage did not significantly differ between conditions ($\beta$ = +0.033, 95\% CI [$-$0.032, 0.099], z = 1.00, p = .317). Taken together, these results provide partial support for H2: the system improved the evenness of perspectives expressed by reducing the dominance of a few perspectives, though it did not increase the total number of distinct viewpoints expressed.

\subsubsection{Argument Strength}
We next examined argument strength using the Supported-Claim Ratio (SCR), defined as the proportion of claims accompanied by at least one supporting premise or piece of evidence. Sentences were labeled (claim, premise, evidence, other) via zero-shot classification with facebook/bart-large-mnli, and SCR was computed at the level of individual participants within conditions. Mixed-effects linear models with participant random intercepts and article variance components indicated a baseline mean of 0.228 and 0.153 under system, yielding a significant decrease ($\beta$ = $-$0.075, SE = 0.027, 95\% CI [$-$0.128, $-$0.021]).
Accordingly, these results do not support H3, indicating that the structured participation did not translate into stronger explicit evidence-based reasoning.

\subsubsection{Emotion}
To examine the effect of the system on emotional expression, we classified sentence-level probabilities for 29 fine-grained emotions. From these outputs, we derived comment-level measures: (a) emotionality (1 $-$ P(neutral)) and (b) probabilities for six core emotions (anger, joy, sadness, fear, disgust, surprise). Mixed-effects logistic models with participant random intercepts were fit, with outcomes logit-transformed. Results indicated a significant reduction in overall emotionality under system support (baseline = 0.656, system = 0.552, $\beta$ = $-$0.109, SE = 0.019, 95\% CI [$-$0.147, $-$0.071], p < .001). Core emotion analyses revealed significant decreases in joy ($\beta$ = $-$0.529, p < .001), sadness ($\beta$ = $-$0.590, p < .001), fear ($\beta$ = $-$0.512, p < .001), and surprise ($\beta$ = $-$0.383, p < .01), while anger and disgust showed small, non-significant increases. Taken together, these findings suggest that system assistance dampened emotional expression across comments, yielding a more neutral and analytical tone. Both positive and negative emotions were suppressed, while antagonistic emotions such as anger and disgust remained unchanged, indicating that the system primarily reduced emotionality without amplifying hostility.
Thus, these results offer partial support for H4: the system discouraged negative affect without promoting positive expression.

\subsubsection{Politeness}
Lastly, to examine effects on politeness, we extracted sentence-level strategies using ConvoKit and grouped them into positive politeness (e.g., please, gratitude, apology) and negative or impolite strategies (e.g., direct address, swearing, negation). For each comment, a politeness score was computed, and participant-level averages were analyzed across conditions using mixed-effects linear models with participant random intercepts and article variance components. Results indicated a baseline mean of 0.573 and a system mean of 0.581 ($\Delta$ = +0.007; $\beta$ = 0.007, SE = 0.009, 95\% CI [$-$0.011, 0.025]), showing no significant difference. These findings suggest that politeness levels remained stable, with the system neither enhancing nor diminishing prosocial tone, thereby preserving civility across conditions.
Accordingly, these results do not support H5, indicating that structured participation preserved politeness in discussion.

\subsection{RQ3. How does the system support users’ experiences of participating in online discussions?}
The post-survey results showed that participants responded positively to the system’s ability to improve their understanding of issues, discussion flow, and diverse viewpoints. Guided topics and questions were particularly effective in helping participants identify key aspects of the articles, stay focused on central points without being distracted by large volumes of comments, and articulate their own thoughts more clearly. Clustering and summarization features provided concise overviews of discussion threads, allowing participants to grasp core ideas quickly, spot gaps, and connect with related opinions. These features also broadened perspectives by grouping similar but differently worded comments and highlighting important flows of discussion, which encouraged participants to think more fluently and from multiple angles. Overall, participants indicated that the system enhanced their comprehension and engagement across three stages of commenting: reading and understanding, structuring and writing, and engaging in meaningful discussion. The following sections present the analyzed results of interviews about how our system supported each stage of this process.

\subsubsection{Reading the Article and Comprehending the Ongoing Discussion}
Understanding both the article and the ongoing discussion is the first step in participating in the discussion. 
Our analysis revealed three key aspects of how the system improved and altered this experience: 
(1) providing access to both high-level overviews and in-depth details to clarify the discussion flow, (2) introducing a bidirectional way of reading between articles and comments to shift through diverse perspectives, and (3) streamlining navigation to help participants focus on relevant points.
All participants noted at least one of these three aspects.

\paragraph{Improved Access to Both High-Level Overviews and In-Depth Information}

Eleven participants (P1–7, P10–13) reported that the system supported efficient navigation of articles and discussions by presenting multiple levels of detail. The summaries first provided quick overviews of representative views and opinion distributions.
\begin{quote}
{It’s really effective for quickly understanding high-level, representative thoughts. The summaries were a great way to see how various opinions are distributed. – P1}
\end{quote}

Clustering and threads supported deeper engagement by breaking down ideas and clarifying what would be discussed under specific questions. Unlike the baseline, which was demanding to read comment by comment, the system encouraged more exploration of individual opinions (P1, P2, P3, P4, P6, P7, P12).

\begin{quote}
{Clustering helped me break things down and go through people’s thoughts and opinions even when I wasn't interested in the topic. – P3}
\end{quote}
\begin{quote}
{With the system, I spent more time reading different opinions. It helped me organize fragmented thoughts, so I ended up dedicating more time to others’ comments. – P4}
\end{quote}

This layered access also supported comprehension of the article, as participants could revisit content with a clearer sense of others’ reasoning.
\begin{quote}
{After reading the comments, going back to the article made it much easier to see where people were coming from. – P13}
\end{quote}

\paragraph{Bidirectional Reading Between Articles and Comments Drives Perspective Shifts}

Six participants (P2, P4, P6, P9, P11, P13) emphasized that the system supported a dynamic, two-way reading flow. Instead of moving linearly from article to comments, they often began with discussion threads to preview key issues before returning to the article with clearer expectations.
\begin{quote}
{Reading the discussion first gave me a sense of what the article would cover. When I went back, the content stuck better and I knew which perspective to take, which I found really helpful. – P4}
\end{quote}

This bidirectional process encouraged perspective shifts and helped participants refine their thoughts while reading.

\begin{quote}
{I could already sort of summarize it in my head by reading the discussion titles first, then reading the article while keeping them in mind. Going backward and afterward, I organized my thoughts based on the topics from the comment section. – P9}
\end{quote}

\begin{quote}
{While reading the whole article, I checked where the threads were formed and how they related to certain points. I kept comparing my thoughts with the thread subjects. – P13}
\end{quote}

\paragraph{Streamlined Comment Navigation for Targeted Focus}

Seven participants (P2, P6, P7, P8, P9, P10, P14) highlighted that the system reduced distractions from scattered or repetitive comments by structuring discussions into threads, summaries, and clusters. This hierarchy offered more targeted navigation than the baseline.
\begin{quote}
{As the number of comments increases, following the conversation becomes challenging within traditional commenting systems. Clustering helped me find the parts I was interested in more effectively. – P7}
\end{quote}

Participants found it easier to locate supporting points and follow coherent topic flows, which improved their ability to focus on relevant aspects of the discussion.

\subsubsection{Structuring Thoughts and Writing Comments}
After understanding the overall issue and discussion content, participants needed to organize their scattered thoughts and engage in writing. While the baseline system often made this process difficult, our system supported users by segmenting disorganized ideas, strengthening arguments through clearer logical direction, and fostering holistic reflection by reminding them of diverse perspectives.

\paragraph{Promoting More Constructive Comments by Segmenting Disorganized Thoughts}

Participants often struggled with scattered ideas covering multiple aspects of an issue in the baseline system. Six participants (P5, P6, P9, P10, P12, P13) noted that our system helped them keep comments specific by separating ideas across clusters. This encouraged focused, single-topic comments and made handling replies easier.
\begin{quote}
{When writing comments, I often wanted to cover many points in one, but since clusters separated them, it was easier to focus on one idea at a time and not include other things as well. – P9}
\end{quote}

By segmenting their thoughts, participants contributed to multiple threads when they had several points to make, resulting in more constructive and organized input. Summarization also prompted reflection on their points and further influenced the development of their thoughts.

\begin{quote}
{The system helped me effectively structure what I wanted to say. Summarization highlighted key points, sometimes pointing out an idea better than I had phrased it, which influenced my thinking and made me realize which points were worth considering.– P6}
\end{quote}

\paragraph{Strengthening Arguments Through Structured Logical Direction}

Five participants (P1, P2, P4, P7, P14) found the system useful for shaping the logical direction of their comments. By showing how stances and arguments were divided, it helped them gather evidence, identify reasoning, and refine their own positions.
\begin{quote}
{After forming an initial stance, I looked at how others’ arguments were divided. Seeing supporting and opposing views helped me organize hints and evidence for my own arguments. – P4}
\end{quote}

Clusters and summarization further supported this process by pinpointing comments with similar and contrasting opinions, making it easier for participants to add their own thoughts with supporting evidence.
\begin{quote}
{Looking at clusters, I easily found people with both similar and different views. This helped me strengthen my arguments by examining the grouped data. – P2}
\end{quote}

\paragraph{Fostering Holistic Reflection of Viewpoints by Reminding Overlooked Perspectives}

Eight participants (P1, P3, P4, P5, P6, P8, P9, P12) noted that lengthy articles often caused them to overlook key points, especially those introduced early. The system reminded them of these missed viewpoints through guided topics and questions.
\begin{quote}
{Articles are long and touch on multiple aspects. I usually only remember the last parts. The system’s topics reminded me of points I had thought about earlier but forgotten. – P9}
\end{quote}

This broader framing encouraged participants to reflect on a wider range of perspectives rather than focusing narrowly on individual comments. Responses indicated that features such as clustering encouraged them to think about the interrelations between different comments (P5), thereby processing the entire comment section together by reflecting on the range of perspectives (P4).

\subsubsection{Building and Contributing in Meaningful Discussions}
Our system improved how participants engaged in discussions by making common ground more visible, helping them identify meaningful opportunities to contribute by bringing like-minded opinions together, and encouraging group-oriented participation. Participants described feeling that their contributions were situated within a more collective process, which reduced barriers to expression and fostered more thoughtful and responsible engagement.

\paragraph{Increased Accessibility for Expressing Opinions by Bringing Together Like-Minded Individuals}

Ten participants (P1, P3, P4, P6, P7, P8, P9, P11, P13, P14) reported that threads and summarization features provided a common ground that made expressing opinions less burdensome. By grouping like-minded perspectives, the system reduced the stress of encountering unexpected counterarguments and made participation feel easier and more connected.
\begin{quote}
{In traditional systems, discussions often end once opposing views appear, and the flow does not last long, making it hard to find opportunities to join in. With categorized topics and maintained direction, the system created an environment where participation was easier. – P7}
\end{quote}
\begin{quote}
{The summarization feature, in a way, gathers people with similar thoughts. It made it easier to express my opinions. – P8}
\end{quote}

Being able to engage with like-minded people improved accessibility compared to the baseline (P3, P6, P8, P9), showing the value of constraints—by designing the space with common ground, the system demonstrated that setting boundaries can help people focus better and engage more effectively in discussions.

\paragraph{Identifying Opportunities to Contribute by Addressing Gaps in the Discussion}

Five participants (P1, P2, P6, P9, P10) noted that clustering made it visually clearer which perspectives were already represented, helping them avoid repetition and instead add missing viewpoints. Unlike the baseline, where it was often unclear whether an idea had already been mentioned, the system highlighted gaps in the discussion and gave users greater inclination to contribute knowledge that had not yet been addressed.
\begin{quote}
{In the baseline, I couldn’t always tell if a point had already been made. Here, it was easier to see what was missing, so I felt more inclined to add new insights. – P9}
\end{quote}

\paragraph{Group-Oriented Contribution with Consideration of Collective Impact}

Through having distributed roles, four participants (P7, P8, P11, P14) described becoming more mindful of their collective impact. Seeing how comments were summarized and grouped prompted them to consider how their input would influence group-oriented contributions, giving them a sense of being grouped with others.
\begin{quote}
{When posting, I thought more about how my comments might impact others. Since summaries reflected shared opinions, I paid more attention to whether my comments fit. – P8}
\end{quote}

By observing how discussions were dynamically shaped through grouping and collective contributions, participants felt that their input would be reviewed, built upon, and incorporated into others’ work. This encouraged them to contribute more thoughtfully and with greater responsibility, offering valuable insights to other participants.

\subsubsection{Challenges in Discussion Participation}

While the system had a positive impact on the overall process of commenting, the post-survey also revealed two limitations: (1) participants sometimes shifted focus toward managing the discussion space due to distributed roles, and (2) the predefined topics, though useful, occasionally felt restrictive. 

During the interview, we asked each participant whether they agreed with these limitations on a five-point scale (Strongly Disagree to Strongly Agree) and in what ways they felt or did not feel these aspects. In this section, we describe the contrasting views on both issues.

\paragraph{Shifting Focus to a Managerial Role, Limiting Participation in Comment Writing}

Participants rated the concern of distributed roles limiting their commenting with a mean of 2.14 ($SD=1.51$) out of 5. Three participants described the role as effortful and sometimes discouraging, with P11 noting hesitation to comment in order to remain objective.

However, most participants felt that roles did not constrain their participation. P4 explained that when lacking expertise, a managerial role was less burdensome and even beneficial, since it allowed them to engage with content more broadly and learn before commenting:

\begin{quote}
{When I have background knowledge, I normally contribute a lot. But when I don’t, it’s challenging to comment. Taking a managerial role helped me explore new perspectives and become more knowledgeable before commenting. – P4}
\end{quote}

\paragraph{Narrowing the Scope of Discussion Due to Limited Topic Range}\label{section:5.6.2}

Participants rated topic limitation concerns at a mean of 2.25 ($SD=1.28$). Several (P1, P4, P7, P8, P9) noted that predefined topics lowered barriers and provided a starting point but risked confining discussion and overlooking other relevant issues.

Others (P3, P5, P6, P10, P14) emphasized that the topics sufficiently covered the range of main points.
Some even noted that without guided topics, they would have felt lost and unable to express their thoughts.
\begin{quote}
{Although there were sub-topics not explicitly suggested, they were related to the main ones, so there was no reason not to comment. – P13}
\end{quote}

\section{Discussion}

Our system demonstrated that fostering collective aspects of user participation can lead to more constructive and meaningful discussions, significantly enhancing the overall quality of discourse. Throughout the study, we found that this improvement was driven by two key factors: (1) the implementation of distributed roles, and (2) the impact of the collective output within a structured discussion space.

Our system's introduction of distributed roles in user participation allowed individuals to actively shape the discussion space, significantly improving the flow and coherence of conversations. 
The collective output from these interactions led to increased engagement, as users were more mindful of how their contributions integrated with the overall conversation. This collaborative approach led to more thoughtful contributions and greater efforts to address and build upon existing viewpoints, thereby enhancing the richness and depth of the discussion.

\subsection{Design Considerations for a Collaborative Approach}
Creating a space where all users collectively work together to create cohesive output involves significant design considerations, as system features and user roles must be thoughtfully aligned and interrelated. While we aimed to address various factors in developing our system, several key takeaways emerged that build upon our design approach.

\subsubsection{Hierarchical Structure and Aggregated Viewpoints}
Our system's hierarchical view of discussion~--~incorporating threads, summarization, clusters, and comments~--~was designed to offer a structured output, enabling users to navigate and streamline their focus efficiently. 
However, it also risked overlooking important details and nuanced individual perspectives. 
For instance, P11 highlighted the significance of precise personal viewpoints in comments, particularly in discussions about serious and important issues, suggesting that summarization might sometimes obscure critical insights if not carefully managed. 
Although participants did not report distortions or unnecessary summarization during the study, the findings underscore the importance of ensuring that collaborative structures make the underlying reasoning not only streamlined but also transparent and recoverable for users who seek deeper engagement.

Additionally, concerns were raised about the potential bias or misinterpretation in the summaries. Although summaries provided a helpful overview, there were questions about how to ensure an objective representation of the discussion. To address this, our system incorporated AI to assist in generating summaries with the intention of reducing personal bias and alleviating the manual burden of summarization. Despite no reported issues of summaries being unnecessary or inaccurately reflecting the discussion points during the study, it remains crucial to carefully design the interaction between users and the system to mitigate these risks.

\subsubsection{Balancing Between Validity and Immediacy}
Designing a collaborative system requires careful consideration of the balance between immediacy and validity. To create collective output that reflects thoughtful deliberation, individual contributions must be validated before being reflected in the system, which introduces a delay and can reduce the immediacy of feedback. This delay means that even if participants contribute effectively, their actions take time to influence the final output.

To ensure validity, our system incorporated a review process within user roles to ensure that individual contributions met quality standards. To minimize the delay between user actions and their visible impact on the system while addressing validity, we balanced factors such as the level of review, the number of reviewers, the required number of users at each level, and the timing of each level's activities based on observations from the pilot study. Despite these efforts, our study found that it took over almost a day for clusters and summaries to appear in the discussion space for all articles. This delay led to some participants feeling that their contributions were not promptly reflected.

Our system showed that balancing immediacy and validity remains a challenging aspect of building a collaborative space, but it is crucial for improving the responsiveness and reliability of the system at the same time.  For example, while our study manually assigned predefined levels for each user, future systems could automatically assign user levels based on the current needs of the discussion space. There is much design space to explore improvements in how these elements are managed to create a more effective and engaging collaborative environment.

\subsubsection{Power structures within Role Hierarchy}
Taking on community roles is a gradual process where active members take on increasing responsibility for management over time~\cite{preece2009reader}. Our design can be interpreted as embedding a power structure into distributed roles, with lower-level users contributing to smaller discussion units and having their activities reviewed by users at the same or higher levels. While participation levels can be an important factor in assigning users to higher roles, they are not the only consideration, as meaningful contributions to broader discussion units often rely on a user's knowledge or understanding of the topic, which may not always directly correlate with their participation levels.

Due to the limitations of our controlled study, we were unable to fully implement the power structure design within user role assignment, as we could not accurately predict the users’ levels of contribution. While we could not factor in users' contribution levels when determining role assignments, in real-world applications, the choice of factors to define the power structure should be carefully considered. Here, we present several design options for structuring the power hierarchy that we've previously considered.

\begin{itemize}
	\item Reputation-based approach: Focuses on the quality of contributions, assigning roles based on metrics such as well-received comments or a high number of positive reactions
	\item Participation-based approach: Evaluates users based on activity thresholds, allowing them to level up by meeting specific engagement criteria
	\item Expertise-driven approach: Prioritizes users' knowledge or interest in a particular topic, assigning roles to those with relevant expertise 
 	\item Community-driven approach: Relies on peer recommendations, where users nominate or endorse others for roles based on their trust and evaluation of their contributions
\end{itemize}

Designing a power structure within role hierarchies requires careful consideration of the system's objectives, the nature of user contributions, and the dynamics of the community. By thoughtfully adapting these methods, systems can create a flexible environment that supports both individual engagement and collective success in managing discussions.

\subsection{Navigating Trade-off Values in Collective Discourse}
In this section, we discuss the trade-offs inherent in the values presented by our system, drawing from our findings. By examining these trade-offs, we explore their implications, clarify the scope of what our system addresses, and identify areas that require additional consideration for future development.

\subsubsection{Building Collective Discourse vs. Workload for Maintaining the Discussion}
Building collective discourse allowed users to explore diverse viewpoints, fostering a deeper understanding of complex issues. However, maintaining such discussions imposes a significant workload on users tasked with organizing and structuring the conversation, which can divert their focus from engagement in commenting. In real-world settings, further efforts should focus on leveraging the collective effort and scale of the community to distribute the workload more effectively. This could involve designing smaller, more manageable tasks, such as tagging or summarizing portions of the discussion. Additionally, systems could actively incentivize users who are willing to take on more responsibility, recognizing that not all users contribute equally. By focusing the workload on those willing and able to contribute more, we can create a more sustainable approach to managing collective discourse.

\subsubsection{Bringing Like-Minded Individuals Together vs. Risk of Echo Chambers}
Our findings showed that the system brought like-minded individuals together, which positively impacted users by helping them connect with others who shared similar perspectives and reducing the burden of expressing their opinions. However, this also raises concerns about the potential risk of echo chambers, as echo chambers can emerge when groups form around shared views, reinforcing their beliefs while excluding opposing opinions~\cite{cinelli2021echo}. Interestingly, our findings indicate that bringing like-minded individuals together did not necessarily lead to reduced diversity or increased polarization, as different steps in the commenting process actively supported both aspects. While the system provided common ground for easier engagement during the contribution phase, it also simultaneously offered overviews of diverse viewpoints and reminders of overlooked perspectives during the reading and structuring phases. By thoughtfully balancing these aspects, we highlight the potential for creating discussion environments that maintain diversity without sacrificing the benefits of shared focus.

\subsubsection{Lowering Barrier of Engagement vs. Preserving Novelty of User Contribution}
Our system utilized AI-generated suggestions for tasks such as summarizing discussions and creating threads. While these features helped lower barriers to engagement, they also present a trade-off by partially delegating the task of discussion framing to the AI, potentially discouraging the introduction of novel ideas or unique contributions, as presented in Section~\ref{section:5.6.2}. To address this challenge, it is crucial to thoughtfully integrate AI within the system flow, ensuring that it serves as a supportive tool rather than a restrictive force. For instance, AI-suggested workflows can be designed as opt-in features, allowing users to control when and how they wish to engage with AI assistance. Alternatively, the system could incorporate a more scaffolded approach to AI support, one that enhances users' ability to express their unique ideas while providing guidance to help present them effectively. We highlight the careful use of AI to preserve users' originality for presenting their unique viewpoints to contribute to online discourse.

\subsection{Implications for Social Dynamics in Online Discussions}
While our system demonstrated the potential of structured collaboration in online news discussion, reorganizing comments fundamentally reshapes the social dynamics through which discourse emerges. We discuss implications in relation to user motivation, social influence patterns, and sustainable participation at scale.

\subsubsection{Supporting Diverse Motivations for Participation}
In this paper, our focus was primarily on demonstrating the user experience and value of a more collaborative commenting system. However, we did not delve deeply into the motivations that drive user participation in news commenting.

Participants' motivations for engaging in discussions are multifaceted, encompassing cognitive, entertainment, social-integrative, and personal identity dimensions~\cite{springer2015user}, and understanding these motivations is crucial for further developing and optimizing collaborative systems. The interview revealed the diverse motivations for participating in discussions, encompassing all of the mentioned four dimensions: engaging in commenting to correct errors or misinformation (cognitive), perceiving commenting as an entertaining activity that adds prestige to the discussion (entertainment), and expressing personal opinions (personal identity).

Future research could investigate how these motivational factors interact with collaborative systems compared to traditional commenting environments. Research could also focus on designing features and collaborative roles that align with these diverse motives, as well as developing strategies to incentivize participation based on users' specific motivations. Throughout the study, participants provided feedback that highlighted their motivations, such as the desire to follow up on comments they enjoyed (P3) and to receive notifications about their activity to see how others reacted to their opinions (P10, P13). Incorporating these factors into our collaborative system through distributed roles will broaden the design space for building a collective discussion space.

\subsubsection{Managing Conformity and Social Influence}
Re-structuring comments also restructures how social cues are revealed, which may shape influence and conformity behaviors. Prior works in social computing demonstrate that individuals often rely on visible popularity cues and majority viewpoints when forming opinions~\cite{wijenayake2020effect, colliander2019fake}. By design, our system counteracts some of these effects by presenting a more evenly weighted distribution of perspectives and reducing early dominance patterns typically reinforced by chronological or popularity-driven feeds. However, aggregated summaries, if not carefully contextualized, risk being perceived as consensus rather than as one representation of the discussion. As a result, while fragmentation and extreme polarization may be reduced, there remains a possibility that conformity pressures become subtler, guiding users toward what appears collectively endorsed and diminishing opportunities for less prominent viewpoints to surface. To ensure that collaboration supports deliberation rather than conformity, future system designs should consider mechanisms that make the diversity and provenance of viewpoints legible, allowing users to appreciate both common ground and unresolved disagreement.

\subsubsection{Incentivizing Knowledge-Building at Scale}
A collaborative system introduces additional coordination effort, which becomes particularly challenging in real-world situations where discussions unfold rapidly or reach large scale. As open collaboration scales, increased governance needs can burden or discourage newcomers \cite{halfaker2013rise}, and the organizational labor that sustains communities often remains invisible and undervalued \cite{matias2019civic}, challenging the long-term sustainability of user-driven knowledge building. Participants in our study welcomed the sense of agency associated with curating and organizing discourse, yet in settings with hundreds or thousands of comments especially during breaking news events, the cognitive burden of coordination could outweigh the perceived benefits. The sustainability of such systems thus depends not only on interaction design but also on how they support meaningful labor. This may involve intelligent automation that reduces repetitive tasks, lightweight contribution opportunities that allow users to help without extensive time investment, and incentive structures that recognize and reinforce knowledge-building behaviors over passive or reactive participation. Rather than minimizing effort entirely, the goal is to align users’ invested effort with visible improvements in collective understanding and discourse quality.

\subsection{Limitations}
While our study provides valuable insights into designing collaborative commenting systems, several limitations should be acknowledged to contextualize our findings and guide future work.

\subsubsection{Methodological Scale and Sampling Constraints}
First, the study involved a relatively small number of participants (N=38) interacting within a controlled discussion setting. In larger discussion spaces, uncivil behaviors can become more prevalent and problematic, impacting a broader audience. Future work should evaluate the system’s performance at scale to understand its robustness in maintaining coherence within extended and high-velocity threads.

Additionally, the study was conducted over a comparatively short duration~--~three days per condition with one article per day. Although this design reflects common attention patterns in real-world news consumption~\cite{kaltenbrunner2007homogeneous}, longer-term deployments are needed to observe how participation evolves over sustained engagement cycles and whether collaboration dynamics persist beyond initial novelty.

The demographic composition of our participants also limits generalizability. Most users were university students with high digital literacy, whereas real-world platforms often host a broader range of ages, backgrounds, and commenting norms.
Moreover, as all interactions occurred in English, varying levels of language proficiency may have shaped how comfortably participants articulated arguments or engaged in coordination.
Future studies should include more diverse populations and content sources to capture a wider variety of interaction styles, language abilities, and topic domains.

\subsubsection{Limited Ecological Representation of Role Hierarchies}
To create a controlled environment for examining role-based collaboration, users were assigned fixed levels rather than earning roles dynamically. While this allowed systematic observation of how roles influence participation, it does not fully represent how power structures form and shift in real-world communities. Alternative designs could allow user levels to be adjusted dynamically, such as through a level-up system based on the completion of specific number of activities or by aligning with user preferences. Longer engagements could incorporate adaptive role mechanisms to better reflect evolving power structures and authority within online discourse.

\subsubsection{Design Limitations in Supporting Constructive Depth and Novelty}
Despite improvements in perspective balance and user participation, the system showed limitations in supporting explicitly reasoned and novelty of contributions. The observed reduction in Supported-Claim Ratio (SCR) suggests that, while users remained engaged, their comments contained fewer explicit justifications. One plausible interpretation is that the structured elements of the system~--~such as guided prompts and cluster-level summaries~--~captured much of the shared reasoning, reducing the need for users to restate evidence at the level of individual comments. Although this may reduce redundancy, it can obscure the transparency and traceability of argumentative depth. Ensuring that collective structures remain accountable to the arguments they represent will therefore require interface mechanisms that make the underlying rationale easily discoverable when needed.

Additionally, although the system successfully prevented the dominance of majority viewpoints by balancing visibility (H\_norm), it did not significantly expand the overall range of distinct viewpoints (Coverage). This suggests that collaboration may help users converge on and refine existing ideas rather than encouraging the emergence of entirely new ones. While convergence can be beneficial for coherence and shared understanding, constructive discourse also relies on the introduction of novel or previously underrepresented insights. Designing features that support viewpoint evolution~--~such as prompts for alternative reasoning, mechanisms for merging across clusters, or identifying missing angles~--~may help prevent discussions from stabilizing prematurely around familiar views.

\section{Conclusion}

In this paper, we explored the design of a commenting system for news outlets, aiming to foster collective aspects of user participation to create a more constructive and meaningful discussion space. By implementing the concept of ``distributed roles'' within the discussion space, our system aimed to enrich discussions by incorporating diverse perspectives while also fostering shared responsibilities in contributions. We designed our system with three core features~--~clusters, summarization, and threads~--~each of which was implemented through roles assigned at three different levels of users. 
The user study with 38 participants showed increased engagement in commenting, with comments demonstrating brevity while maintaining analytical complexity and a reduction in emotional expression. The findings from 14 follow-up interviews suggest that the system positively impacted various phases of the comment writing experience, including reading articles, structuring thoughts, and contributing to discussions. Our results indicate that the system effectively built a structured and organized space, promoting more thoughtful and constructive comment writing through collective behavior. We conclude by highlighting key design considerations and trade-offs in our system, offering guidance for the development of future discourse systems using a collaborative approach. Future research could explore various approaches to designing distributed user levels and roles, including methods for embedding power structures within role assignments and strategies for task assignment that adapt to community size.

\begin{acks}
This research was supported by the KAIST New Faculty Research Fund (Project No. G04230040: Building Integrated Models for Online Governance).
\end{acks}

\bibliographystyle{ACM-Reference-Format}
\bibliography{main}

@article{elo2008qualitative,
author = {Elo, Satu and Kyngäs, Helvi},
title = {The qualitative content analysis process},
journal = {Journal of Advanced Nursing},
volume = {62},
number = {1},
pages = {107-115},
doi = {https://doi.org/10.1111/j.1365-2648.2007.04569.x},
url = {https://onlinelibrary.wiley.com/doi/abs/10.1111/j.1365-2648.2007.04569.x},
eprint = {https://onlinelibrary.wiley.com/doi/pdf/10.1111/j.1365-2648.2007.04569.x},
year = {2008}
}

@article{springer2015user,
title={User comments: Motives and inhibitors to write and read},
author={Springer, Nina and Engelmann, Ines and Pfaffinger, Christian},
journal={Information, Communication \& Society},
volume={18},
number={7},
pages={798--815},
year={2015},
publisher={Taylor \& Francis}
}

@article{tian2021system,
  title={A system for interleaving discussion and summarization in online collaboration},
  author={Tian, Sunny and Zhang, Amy X and Karger, David},
  journal={Proceedings of the ACM on Human-Computer Interaction},
  volume={4},
  number={CSCW3},
  pages={1--27},
  year={2021},
  publisher={ACM New York, NY, USA}
}

@inproceedings{zhang2017wikum,
  title={Wikum: Bridging discussion forums and wikis using recursive summarization},
  author={Zhang, Amy X and Verou, Lea and Karger, David},
  booktitle={Proceedings of the 2017 ACM Conference on Computer Supported Cooperative Work and Social Computing},
  pages={2082--2096},
  year={2017}
}

@inproceedings{kriplean2012you,
  title={Is this what you meant? Promoting listening on the web with reflect},
  author={Kriplean, Travis and Toomim, Michael and Morgan, Jonathan and Borning, Alan and Ko, Amy J},
  booktitle={Proceedings of the SIGCHI Conference on Human Factors in Computing Systems},
  pages={1559--1568},
  year={2012}
}

@inproceedings{farnham2000structured,
  title={Structured online interactions: improving the decision-making of small discussion groups},
  author={Farnham, Shelly and Chesley, Harry R and McGhee, Debbie E and Kawal, Reena and Landau, Jennifer},
  booktitle={Proceedings of the 2000 ACM conference on Computer supported cooperative work},
  pages={299--308},
  year={2000}
}

@inproceedings{nam2007arkose,
  title={Arkose: reusing informal information from online discussions},
  author={Nam, Kevin K and Ackerman, Mark S},
  booktitle={Proceedings of the 2007 ACM International Conference on Supporting Group Work},
  pages={137--146},
  year={2007}
}

@inproceedings{rambow2004summarizing,
  title={Summarizing email threads},
  author={Rambow, Owen and Shrestha, Lokesh and Chen, John and Laurdisen, Christy},
  booktitle={Proceedings of HLT-NAACL 2004: Short Papers},
  pages={105--108},
  year={2004}
}

@article{kim2021moderator,
  title={Moderator chatbot for deliberative discussion: Effects of discussion structure and discussant facilitation},
  author={Kim, Soomin and Eun, Jinsu and Seering, Joseph and Lee, Joonhwan},
  journal={Proceedings of the ACM on Human-Computer Interaction},
  volume={5},
  number={CSCW1},
  pages={1--26},
  year={2021},
  publisher={ACM New York, NY, USA}
}

@inproceedings{lee2020solutionchat,
  title={Solutionchat: Real-time moderator support for chat-based structured discussion},
  author={Lee, Sung-Chul and Song, Jaeyoon and Ko, Eun-Young and Park, Seongho and Kim, Jihee and Kim, Juho},
  booktitle={Proceedings of the 2020 CHI conference on human factors in computing systems},
  pages={1--12},
  year={2020}
}

@inproceedings{munson2013encouraging,
  title={Encouraging reading of diverse political viewpoints with a browser widget},
  author={Munson, Sean and Lee, Stephanie and Resnick, Paul},
  booktitle={Proceedings of the international AAAI conference on web and social media},
  volume={7},
  number={1},
  pages={419--428},
  year={2013}
}

@inproceedings{kriplean2012supporting,
  title={Supporting reflective public thought with considerit},
  author={Kriplean, Travis and Morgan, Jonathan and Freelon, Deen and Borning, Alan and Bennett, Lance},
  booktitle={Proceedings of the ACM 2012 conference on Computer Supported Cooperative Work},
  pages={265--274},
  year={2012}
}

@inproceedings{gao2018burst,
  title={Burst your bubble! an intelligent system for improving awareness of diverse social opinions},
  author={Gao, Mingkun and Do, Hyo Jin and Fu, Wai-Tat},
  booktitle={Proceedings of the 23rd International Conference on Intelligent User Interfaces},
  pages={371--383},
  year={2018}
}

@article{nelimarkka2019re,
  title={(Re) Design to Mitigate Political Polarization: Reflecting Habermas' ideal communication space in the United States of America and Finland},
  author={Nelimarkka, Matti and Rancy, Jean Philippe and Grygiel, Jennifer and Semaan, Bryan},
  journal={Proceedings of the ACM on Human-computer Interaction},
  volume={3},
  number={CSCW},
  pages={1--25},
  year={2019},
  publisher={ACM New York, NY, USA}
}

@inproceedings{faridani2010opinion,
  title={Opinion space: a scalable tool for browsing online comments},
  author={Faridani, Siamak and Bitton, Ephrat and Ryokai, Kimiko and Goldberg, Ken},
  booktitle={Proceedings of the SIGCHI Conference on Human Factors in Computing Systems},
  pages={1175--1184},
  year={2010}
}

@book{singer2011participatory,
  title={Participatory journalism},
  author={Singer, Jane B and Reich, Zvi},
  year={2011},
  publisher={Wiley Online Library}
}

@article{lewis2014reciprocal,
  title={Reciprocal journalism: A concept of mutual exchange between journalists and audiences},
  author={Lewis, Seth C and Holton, Avery E and Coddington, Mark},
  journal={Journalism practice},
  volume={8},
  number={2},
  pages={229--241},
  year={2014},
  publisher={Taylor \& Francis}
}

@article{ruiz2011public,
  title={Public sphere 2.0? The democratic qualities of citizen debates in online newspapers},
  author={Ruiz, Carlos and Domingo, David and Mic{\'o}, Josep Llu{\'\i}s and D{\'\i}az-Noci, Javier and Meso, Koldo and Masip, Pere},
  journal={The International journal of press/politics},
  volume={16},
  number={4},
  pages={463--487},
  year={2011},
  publisher={SAGE Publications Sage CA: Los Angeles, CA}
}

@article{santana2011online,
  title={Online readers' comments represent new opinion pipeline},
  author={Santana, Arthur D},
  journal={Newspaper research journal},
  volume={32},
  number={3},
  pages={66--81},
  year={2011},
  publisher={SAGE Publications Sage CA: Los Angeles, CA}
}

@inproceedings{diakopoulos2011towards,
  title={Towards quality discourse in online news comments},
  author={Diakopoulos, Nicholas and Naaman, Mor},
  booktitle={Proceedings of the ACM 2011 conference on Computer supported cooperative work},
  pages={133--142},
  year={2011}
}

@article{reich2011user,
  title={User comments: The transformation of participatory space},
  author={Reich, Zvi},
  journal={Participatory journalism: Guarding open gates at online newspapers},
  pages={96--117},
  year={2011},
  publisher={Wiley Online Library}
}

@article{robinson2010traditionalists,
  title={Traditionalists vs. convergers: Textual privilege, boundary work, and the journalist—Audience relationship in the commenting policies of online news sites},
  author={Robinson, Sue},
  journal={Convergence},
  volume={16},
  number={1},
  pages={125--143},
  year={2010},
  publisher={Sage Publications Sage UK: London, England}
}

@article{quandt2018dark,
  title={Dark participation},
  author={Quandt, Thorsten},
  journal={Media and communication},
  volume={6},
  number={4},
  pages={36--48},
  year={2018},
  publisher={MISC}
}

@article{muddiman2017news,
  title={News values, cognitive biases, and partisan incivility in comment sections},
  author={Muddiman, Ashley and Stroud, Natalie Jomini},
  journal={Journal of communication},
  volume={67},
  number={4},
  pages={586--609},
  year={2017},
  publisher={Oxford University Press}
}

@article{wolfgang2021taming,
  title={Taming the ‘trolls’: How journalists negotiate the boundaries of journalism and online comments},
  author={Wolfgang, J David},
  journal={Journalism},
  volume={22},
  number={1},
  pages={139--156},
  year={2021},
  publisher={SAGE Publications Sage UK: London, England}
}

@article{o2003reconceptualizing,
  title={Reconceptualizing ‘flaming’and other problematic messages},
  author={O’sullivan, Patrick B and Flanagin, Andrew J},
  journal={New media \& society},
  volume={5},
  number={1},
  pages={69--94},
  year={2003},
  publisher={Sage Publications Sage CA: Thousand Oaks, CA}
}

@article{erjavec2012you,
  title={“You don't understand, this is a new war!” Analysis of hate speech in news web sites' comments},
  author={Erjavec, Karmen and Kova{\v{c}}i{\v{c}}, Melita Poler},
  journal={Mass Communication and Society},
  volume={15},
  number={6},
  pages={899--920},
  year={2012},
  publisher={Taylor \& Francis}
}

@article{hughey2013racist,
  title={Racist comments at online news sites: a methodological dilemma for discourse analysis},
  author={Hughey, Matthew W and Daniels, Jessie},
  journal={Media, Culture \& Society},
  volume={35},
  number={3},
  pages={332--347},
  year={2013},
  publisher={SAGE Publications Sage UK: London, England}
}

@inproceedings{lampe2007follow,
  title={Follow the reader: filtering comments on slashdot},
  author={Lampe, Cliff AC and Johnston, Erik and Resnick, Paul},
  booktitle={Proceedings of the SIGCHI conference on Human factors in computing systems},
  pages={1253--1262},
  year={2007}
}

@inproceedings{lampe2004slash,
  title={Slash (dot) and burn: distributed moderation in a large online conversation space},
  author={Lampe, Cliff and Resnick, Paul},
  booktitle={Proceedings of the SIGCHI conference on Human factors in computing systems},
  pages={543--550},
  year={2004}
}

@book{habermas1991structural,
  title={The structural transformation of the public sphere: An inquiry into a category of bourgeois society},
  author={Habermas, Jurgen},
  year={1991},
  publisher={MIT press}
}

@book{dewey2012public,
  title={The public and its problems: An essay in political inquiry},
  author={Dewey, John and Rogers, Melvin L},
  year={2012},
  publisher={Penn State Press}
}

@article{dahlberg2007internet,
  title={The Internet, deliberative democracy, and power: Radicalizing the public sphere},
  author={Dahlberg, Lincoln},
  journal={International journal of media \& cultural politics},
  volume={3},
  number={1},
  pages={47--64},
  year={2007},
  publisher={Intellect}
}

@article{hsueh2015leave,
  title={“Leave your comment below”: Can biased online comments influence our own prejudicial attitudes and behaviors?},
  author={Hsueh, Mark and Yogeeswaran, Kumar and Malinen, Sanna},
  journal={Human communication research},
  volume={41},
  number={4},
  pages={557--576},
  year={2015},
  publisher={Oxford University Press Oxford, UK}
}

@article{nelson2021killing,
  title={Killing the comments: Why do news organizations remove user commentary functions?},
  author={Nelson, Maria N and Ksiazek, Thomas B and Springer, Nina},
  journal={Journalism and Media},
  volume={2},
  number={4},
  pages={572--583},
  year={2021},
  publisher={MDPI}
}

@inproceedings{seering2019designing,
  title={Designing user interface elements to improve the quality and civility of discourse in online commenting behaviors},
  author={Seering, Joseph and Fang, Tianmi and Damasco, Luca and Chen, Mianhong'Cherie' and Sun, Likang and Kaufman, Geoff},
  booktitle={Proceedings of the 2019 CHI conference on human factors in computing systems},
  pages={1--14},
  year={2019}
}

@inproceedings{kiskola2021applying,
  title={Applying critical voice in design of user interfaces for supporting self-reflection and emotion regulation in online news commenting},
  author={Kiskola, Joel and Olsson, Thomas and V{\"a}{\"a}t{\"a}j{\"a}, Heli and H. Syrj{\"a}m{\"a}ki, Aleksi and Rantasila, Anna and Isokoski, Poika and Ilves, Mirja and Surakka, Veikko},
  booktitle={Proceedings of the 2021 CHI conference on human factors in computing systems},
  pages={1--13},
  year={2021}
}

@article{coe2014online,
  title={Online and uncivil? Patterns and determinants of incivility in newspaper website comments},
  author={Coe, Kevin and Kenski, Kate and Rains, Stephen A},
  journal={Journal of communication},
  volume={64},
  number={4},
  pages={658--679},
  year={2014},
  publisher={Oxford University Press}
}

@article{dahlberg2007rethinking,
  title={Rethinking the fragmentation of the cyberpublic: from consensus to contestation},
  author={Dahlberg, Lincoln},
  journal={New media \& society},
  volume={9},
  number={5},
  pages={827--847},
  year={2007},
  publisher={Sage Publications Sage UK: London, England}
}

@article{anderson2014nasty,
  title={The “nasty effect:” Online incivility and risk perceptions of emerging technologies},
  author={Anderson, Ashley A and Brossard, Dominique and Scheufele, Dietram A and Xenos, Michael A and Ladwig, Peter},
  journal={Journal of computer-mediated communication},
  volume={19},
  number={3},
  pages={373--387},
  year={2014},
  publisher={Oxford University Press Oxford, UK}
}

@article{ksiazek2015discussing,
  title={Discussing the news: Civility and hostility in user comments},
  author={Ksiazek, Thomas B and Peer, Limor and Zivic, Andrew},
  journal={Digital journalism},
  volume={3},
  number={6},
  pages={850--870},
  year={2015},
  publisher={Taylor \& Francis}
}

@article{paulussen2011inside,
  title={Inside the Newsroom: Journalists' motivations and organizational structures},
  author={Paulussen, Steve},
  journal={Participatory journalism: Guarding open gates at online newspapers},
  pages={57--75},
  year={2011},
  publisher={Wiley Online Library}
}

@article{mcelroy2013old,
  title={Where old (gatekeepers) meets new (media) Herding reader comments into print},
  author={McElroy, Kathleen},
  journal={Journalism Practice},
  volume={7},
  number={6},
  pages={755--771},
  year={2013},
  publisher={Taylor \& Francis}
}

@article{reimer2023content,
  title={Content analyses of user comments in journalism: A systematic literature review spanning communication studies and computer science},
  author={Reimer, Julius and H{\"a}ring, Marlo and Loosen, Wiebke and Maalej, Walid and Merten, Lisa},
  journal={Digital Journalism},
  volume={11},
  number={7},
  pages={1328--1352},
  year={2023},
  publisher={Taylor \& Francis}
}

@article{kim2021starrythoughts,
  title={StarryThoughts: facilitating diverse opinion exploration on social issues},
  author={Kim, Hyunwoo and Kim, Haesoo and Jo, Kyung Je and Kim, Juho},
  journal={Proceedings of the ACM on Human-Computer Interaction},
  volume={5},
  number={CSCW1},
  pages={1--29},
  year={2021},
  publisher={ACM New York, NY, USA}
}

@book{kraut2012building,
  title={Building successful online communities: Evidence-based social design},
  author={Kraut, Robert E and Resnick, Paul},
  year={2012},
  publisher={Mit Press}
}

@inproceedings{mcinnis2016one,
  title={One and Done: Factors affecting one-time contributors to ad-hoc online communities},
  author={McInnis, Brian James and Murnane, Elizabeth Lindley and Epstein, Dmitry and Cosley, Dan and Leshed, Gilly},
  booktitle={Proceedings of the 19th ACM Conference on Computer-Supported Cooperative Work \& Social Computing},
  pages={609--623},
  year={2016}
}

@inproceedings{mcinnis2018effects,
  title={Effects of comment curation and opposition on coherence in online policy discussion},
  author={McInnis, Brian and Cosley, Dan and Baumer, Eric and Leshed, Gilly},
  booktitle={Proceedings of the 2018 ACM International Conference on Supporting Group Work},
  pages={347--358},
  year={2018}
}

@article{chang2022thread,
  title={Thread with caution: Proactively helping users assess and deescalate tension in their online discussions},
  author={Chang, Jonathan P and Schluger, Charlotte and Danescu-Niculescu-Mizil, Cristian},
  journal={Proceedings of the ACM on Human-Computer Interaction},
  volume={6},
  number={CSCW2},
  pages={1--37},
  year={2022},
  publisher={ACM New York, NY, USA}
}

@article{schluger2022proactive,
  title={Proactive moderation of online discussions: Existing practices and the potential for algorithmic support},
  author={Schluger, Charlotte and Chang, Jonathan P and Danescu-Niculescu-Mizil, Cristian and Levy, Karen},
  journal={Proceedings of the ACM on Human-Computer Interaction},
  volume={6},
  number={CSCW2},
  pages={1--27},
  year={2022},
  publisher={ACM New York, NY, USA}
}

@article{zhang2017crowd,
  title={Crowd development: The interplay between crowd evaluation and collaborative dynamics in wikipedia},
  author={Zhang, Ark Fangzhou and Livneh, Danielle and Budak, Ceren and Robert Jr, Lionel P and Romero, Daniel M},
  journal={Proceedings of the ACM on Human-Computer Interaction},
  volume={1},
  number={CSCW},
  pages={1--21},
  year={2017},
  publisher={ACM New York, NY, USA}
}

@article{im2018deliberation,
  title={Deliberation and resolution on wikipedia: A case study of requests for comments},
  author={Im, Jane and Zhang, Amy X and Schilling, Christopher J and Karger, David},
  journal={Proceedings of the ACM on Human-Computer Interaction},
  volume={2},
  number={CSCW},
  pages={1--24},
  year={2018},
  publisher={ACM New York, NY, USA}
}

@inproceedings{hoque2015convisit,
  title={Convisit: Interactive topic modeling for exploring asynchronous online conversations},
  author={Hoque, Enamul and Carenini, Giuseppe},
  booktitle={Proceedings of the 20th International Conference on Intelligent User Interfaces},
  pages={169--180},
  year={2015}
}

@inproceedings{hoque2016multiconvis,
  title={Multiconvis: A visual text analytics system for exploring a collection of online conversations},
  author={Hoque, Enamul and Carenini, Giuseppe},
  booktitle={Proceedings of the 21st international conference on intelligent user interfaces},
  pages={96--107},
  year={2016}
}

@article{budak2017threading,
  title={Threading is sticky: How threaded conversations promote comment system user retention},
  author={Budak, Ceren and Garrett, R Kelly and Resnick, Paul and Kamin, Julia},
  journal={Proceedings of the ACM on Human-Computer Interaction},
  volume={1},
  number={CSCW},
  pages={1--20},
  year={2017},
  publisher={ACM New York, NY, USA}
}

@article{zhang2018making,
  title={Making sense of group chat through collaborative tagging and summarization},
  author={Zhang, Amy X and Cranshaw, Justin},
  journal={Proceedings of the ACM on Human-Computer Interaction},
  volume={2},
  number={CSCW},
  pages={1--27},
  year={2018},
  publisher={ACM New York, NY, USA}
}

@inproceedings{schneider2011understanding,
  title={Understanding and improving Wikipedia article discussion spaces},
  author={Schneider, Jodi and Passant, Alexandre and Breslin, John G},
  booktitle={Proceedings of the 2011 ACM Symposium on Applied Computing},
  pages={808--813},
  year={2011}
}

@article{wang2022designing,
  title={Designing for engaging with news using moral framing towards bridging ideological divides},
  author={Wang, Jessica Z and Zhang, Amy X and Karger, David R},
  journal={Proceedings of the ACM on Human-Computer Interaction},
  volume={6},
  number={GROUP},
  pages={1--23},
  year={2022},
  publisher={ACM New York, NY, USA}
}

@article{cinelli2021echo,
  title={The echo chamber effect on social media},
  author={Cinelli, Matteo and De Francisci Morales, Gianmarco and Galeazzi, Alessandro and Quattrociocchi, Walter and Starnini, Michele},
  journal={Proceedings of the National Academy of Sciences},
  volume={118},
  number={9},
  pages={e2023301118},
  year={2021},
  publisher={National Academy of Sciences}
}

@article{kaltenbrunner2007homogeneous,
  title={Homogeneous temporal activity patterns in a large online communication space},
  author={Kaltenbrunner, Andreas and G{\'o}mez, Vicen{\c{c}} and Moghnieh, Ayman and Meza, Rodrigo and Blat, Josep and L{\'o}pez, Vicente},
  journal={arXiv preprint arXiv:0708.1579},
  year={2007}
}

@article{preece2009reader,
  title={The reader-to-leader framework: Motivating technology-mediated social participation},
  author={Preece, Jennifer and Shneiderman, Ben},
  journal={AIS transactions on human-computer interaction},
  volume={1},
  number={1},
  pages={13--32},
  year={2009}
}

@article{seering2019moderator,
  title={Moderator engagement and community development in the age of algorithms},
  author={Seering, Joseph and Wang, Tony and Yoon, Jina and Kaufman, Geoff},
  journal={New media \& society},
  volume={21},
  number={7},
  pages={1417--1443},
  year={2019},
  publisher={SAGE Publications Sage UK: London, England}
}

@book{roberts2019behind,
  title={Behind the screen},
  author={Roberts, Sarah T},
  year={2019},
  publisher={Yale University Press}
}

@article{matias2019preventing,
  title={Preventing harassment and increasing group participation through social norms in 2,190 online science discussions},
  author={Matias, J Nathan},
  journal={Proceedings of the National Academy of Sciences},
  volume={116},
  number={20},
  pages={9785--9789},
  year={2019},
  publisher={National Acad Sciences}
}

@article{tenenboim2022comments,
  title={Comments, shares, or likes: What makes news posts engaging in different ways},
  author={Tenenboim, Ori},
  journal={Social Media+ Society},
  volume={8},
  number={4},
  pages={20563051221130282},
  year={2022},
  publisher={SAGE Publications Sage UK: London, England}
}

@inproceedings{munson2010presenting,
  title={Presenting diverse political opinions: how and how much},
  author={Munson, Sean A and Resnick, Paul},
  booktitle={Proceedings of the SIGCHI conference on human factors in computing systems},
  pages={1457--1466},
  year={2010}
}

@article{stromer2007measuring,
  title={Measuring deliberation’s content: A coding scheme},
  author={Stromer-Galley, Jennifer},
  journal={Journal of Deliberative Democracy},
  volume={3},
  number={1},
  year={2007},
  publisher={University of Westminster Press}
}

@article{danescu2013computational,
  title={A computational approach to politeness with application to social factors},
  author={Danescu-Niculescu-Mizil, Cristian and Sudhof, Moritz and Jurafsky, Dan and Leskovec, Jure and Potts, Christopher},
  journal={arXiv preprint arXiv:1306.6078},
  year={2013}
}

@article{chmiel2011negative,
  title={Negative emotions boost user activity at BBC forum},
  author={Chmiel, Anna and Sobkowicz, Pawel and Sienkiewicz, Julian and Paltoglou, Georgios and Buckley, Kevan and Thelwall, Mike and Ho{\l}yst, Janusz A},
  journal={Physica A: statistical mechanics and its applications},
  volume={390},
  number={16},
  pages={2936--2944},
  year={2011},
  publisher={Elsevier}
}

@article{wijenayake2020effect,
  title={Effect of conformity on perceived trustworthiness of news in social media},
  author={Wijenayake, Senuri and Hettiachchi, Danula and Hosio, Simo and Kostakos, Vassilis and Goncalves, Jorge},
  journal={IEEE Internet Computing},
  volume={25},
  number={1},
  pages={12--19},
  year={2020},
  publisher={IEEE}
}

@article{colliander2019fake,
  title={“This is fake news”: Investigating the role of conformity to other users’ views when commenting on and spreading disinformation in social media},
  author={Colliander, Jonas},
  journal={Computers in Human Behavior},
  volume={97},
  pages={202--215},
  year={2019},
  publisher={Elsevier}
}

@article{halfaker2013rise,
  title={The rise and decline of an open collaboration system: How Wikipedia’s reaction to popularity is causing its decline},
  author={Halfaker, Aaron and Geiger, R Stuart and Morgan, Jonathan T and Riedl, John},
  journal={American behavioral scientist},
  volume={57},
  number={5},
  pages={664--688},
  year={2013},
  publisher={SAGE Publications Sage CA: Los Angeles, CA}
}

@article{matias2019civic,
  title={The civic labor of volunteer moderators online},
  author={Matias, J Nathan},
  journal={Social Media+ Society},
  volume={5},
  number={2},
  pages={2056305119836778},
  year={2019},
  publisher={SAGE Publications Sage UK: London, England}
}

\clearpage
\appendix
\begin{appendices}
\appendix

\newpage
\section{Prompts Used in the System}
\label{appendix:prompt}

\subsubsection{Suggested Summarization}
\begin{quote}
\texttt{You are a helpful assistant that summarizes comments from a neutral perspective. Please summarize the following comments from multiple users from the third perspective while paraphrasing bad words, provide a general overview of what the comment thread is saying, and limit the summary to 20 words:~\{~comments~\} Summary:}
\end{quote}

\subsubsection{Suggested Thread Topics and Guiding Questions}
\begin{quote}
    \texttt{You are a helpful assistant that generates topics and questions based on given text. Please generate 4 diverse and distinct topics based on the following article text. For each topic, also generate a thought-provoking question that can open a meaningful conversation among readers and help explore the topic further. Each topic should be represented by a minimum of 4 words and a maximum of 5 words. Format the output as follows: \\
    Topic 1: <topic> \\
    Question 1: <question> \\
    Topic 2: <topic> \\
    Question 2: <question> \\
    Topic 3: <topic> \\
    Question 3: <question> \\
    Topic 4: <topic> \\
    Question 4: <question> \\\\
    Article text: \{~text~\}}
\end{quote}

\section{Post-survey Questions}
\label{appendix:post-survey}

\begin{enumerate}[itemsep=1ex]
    \item How frequently did you visit our system? (Please specify the \textbf{average number of times per day})
    \item How did \textbf{performing the assigned roles} influence your experience in writing comments? 
    \item How did \textbf{having a guided discussion (with discussion topics and guiding questions)} influence your experience in writing comments?
    \item How did \textbf{having clustered and summarized comments} influence your experience in writing comments?
    \item Did the system help you understand the \textbf{ongoing discussion flow}? \textit{(1-Strongly Disagree, 5-Strongly Agree)}
    \begin{enumerate}
        \item How did it help or not help you understand the discussion flow?
    \end{enumerate}
    \item Did the system help you understand \textbf{other people's perspectives on the discussion topic}? \textit{(1-Strongly Disagree, 5-Strongly Agree)}
    \begin{enumerate} 
        \item How did it help or not help you understand other people's perspectives?
    \end{enumerate}
    \item Did the system help you \textbf{understand the issue discussed in the article}? \textit{(1-Strongly Disagree, 5-Strongly Agree)}
    \begin{enumerate}
        \item How did it help or not help you understand the issue discussed in the article?
    \end{enumerate}
\end{enumerate}

\section{Interview Questions}
\label{appendix:interview}

\begin{enumerate}[itemsep=1ex]
    \item \textbf{Comparison of Commenting Experience: Our System vs. Baseline}
    \begin{enumerate}
        \item Can you describe how you used the first system?
        \item Can you describe how you used the second system?
        \item How did your experience differ in terms of reading articles, reviewing others' comments, and writing comments?
    \end{enumerate}
    \par\medskip
    \item \textbf{Impact of the System on Deliberation Experience}
    \begin{enumerate}
        \item \textbf{Accessing Information}: How did the system affect your ability to find and utilize relevant information?
        \item \textbf{Structuring Thoughts} How did the system assist in organizing your thoughts?
        \item \textbf{Engaging in Discussions} How did the system impact your ability to participate in and contribute to discussions?
    \end{enumerate}
    \par\medskip
    \item \textbf{Key Areas of Usefulness of the System}
    \begin{enumerate}
        \item \textbf{Being aware of diverse perspectives when expressing thoughts}
        \begin{enumerate}
            \item For each of the following areas, please rate your level of agreement on a scale of 1 to 5 \textit{(1 = Strongly Disagree, 5 = Strongly Agree)}
            \item Additionally, could you share any specific experiences where you noticed these aspects while using the system? Please provide examples if applicable.
        \end{enumerate}
        \item \textbf{Participating and contributing to more collective actions}
        \begin{enumerate}
            \item For each of the following areas, please rate your level of agreement on a scale of 1 to 5 \textit{(1 = Strongly Disagree, 5 = Strongly Agree)}
            \item Additionally, could you share any specific experiences where you noticed these aspects while using the system? Please provide examples if applicable.
        \end{enumerate}
        \item \textbf{Developing a focused understanding of articles and discussions}
        \begin{enumerate}
            \item For each of the following areas, please rate your level of agreement on a scale of 1 to 5 \textit{(1 = Strongly Disagree, 5 = Strongly Agree)}
            \item Additionally, could you share any specific experiences where you noticed these aspects while using the system? Please provide examples if applicable.
        \end{enumerate}
    \end{enumerate}
    \par\medskip
    \item \textbf{Limitations of the System}
    \begin{enumerate}
        \item \textbf{Shift to managerial Role, limiting participation in commenting}
        \begin{enumerate}
            \item For each of the following areas, please rate your level of agreement on a scale of 1 to 5 \textit{(1 = Strongly Disagree, 5 = Strongly Agree)}
            \item Additionally, could you share any specific experiences where you noticed these aspects while using the system? Please provide examples if applicable.
        \end{enumerate}
        \item \textbf{Limiting the scope of discussion space}
        \begin{enumerate}
            \item For each of the following areas, please rate your level of agreement on a scale of 1 to 5 \textit{(1 = Strongly Disagree, 5 = Strongly Agree)}
            \item Additionally, could you share any specific experiences where you noticed these aspects while using the system? Please provide examples if applicable.
        \end{enumerate}
    \end{enumerate}
    \par\medskip
    \item \textbf{Feedback for Improvement}
    \begin{enumerate}
        \item Do you have any feedback you'd like to provide regarding our system?
    \end{enumerate}
\end{enumerate}

\end{appendices}

\end{document}